\documentclass[10pt]{bmc_article}

\usepackage{cite} 
\usepackage{url}  
\usepackage{ifthen}  
\usepackage{multicol}   
\usepackage{graphicx}
\newboolean{publ}

\newenvironment{bmcformat}{\begin{raggedright}\baselineskip20pt\sloppy\setboolean{publ}{false}}{\end{raggedright}\baselineskip20pt\sloppy}

\begin{document}
\begin{bmcformat}



\title{Parameters of proteome evolution from
histograms of amino-acid sequence identities of paralogous
proteins.}

\author{Jacob Bock Axelsen$^{1,2}$%
       \email{Jacob Bock Axelsen - bock@nbi.dk}%
      \and
         Koon-Kiu Yan$^{2,3}$%
         \email{Koon-Kiu Yan- kyan@bnl.gov}
       and
         Sergei Maslov$^{2,3}$\correspondingauthor%
         \email{Sergei Maslov  - maslov@bnl.gov}%
      }


\address{%
    \iid(1)Center for Models of Life, Niels Bohr Institute, Blegdamsvej 17, DK-2100,
    Copenhagen \O, Denmark\\
    \iid(2)Department of Condensed Matter Physics and Materials Science,
    Brookhaven National Laboratory, Upton, New York 11973, USA\\
    \iid(3)Department of Physics and Astronomy, Stony Brook University, Stony Brook,
    New York 11794, USA
}%

\maketitle


\begin{abstract}
        \paragraph*{Background:} The evolution of the
        full repertoire of proteins encoded in a given
        genome is mostly driven by gene duplications,
        deletions, and sequence modifications of
        existing proteins. Indirect information about
        relative rates and other intrinsic parameters
        of these three basic processes is contained in
        the proteome-wide distribution of sequence
        identities of pairs of paralogous proteins.
        \paragraph*{Results:} We introduce a simple
        mathematical framework based on a stochastic birth-and-death
        model that allows one to
        extract some of this information and apply it
        to the set of all pairs of paralogous proteins
        in {\it H. pylori, E. coli, S. cerevisiae, C.
        elegans, D. melanogaster}, and {\it H. sapiens}.
        It was found that the histogram of sequence identities $p$
        generated by an all-to-all alignment of all
        protein sequences encoded in a genome is well fitted with a power-law
        form $\sim p^{-\gamma}$ with the value of the exponent
        $\gamma$ around 4 for the majority of organisms used in this
        study. This implies that the intra-protein variability of substitution
        rates is best described by the Gamma-distribution with
        the exponent $\alpha \approx 0.33$. Different features of the shape of
        such histograms
        allow us to quantify the ratio between the genome-wide average
        deletion/duplication rates and the amino-acid substitution rate.
        \paragraph*{Conclusions:}
        We separately measure the short-term (``raw'') duplication and deletion rates
        $r_{\mathrm{dup}}^*$, $r_{\mathrm{del}}^*$
        which include gene copies that will be removed soon after
        the duplication event and their dramatically
        reduced long-term counterparts $r_{\mathrm{dup}}$,
        $r_{\mathrm{del}}$.
         High deletion rate among recently duplicated proteins is
         consistent with a scenario in which they didn't have
         enough time to significantly change their
         functional roles and thus are to a large
         degree disposable.
        Systematic trends of each of the four
        duplication/deletion rates with the total number of
        genes in the genome were analyzed.
        All but the deletion rate of recent duplicates
        $r_{\mathrm{del}}^*$ were shown to
        systematically increase with
        $N_{\mathrm{genes}}$.
        Abnormally flat shapes of sequence identity
        histograms observed for yeast and human are consistent with
        lineages leading to these organisms undergoing one or more whole-genome
        duplications. This interpretation is corroborated by
        our analysis of the genome of {\it Paramecium tetraurelia} where
        the $p^{-4}$ profile of the histogram is
        gradually restored by the successive removal of paralogs
        generated in its four known whole-genome duplication events.
        \paragraph*{This article was reviewed by Eugene Koonin,
        Yuri Wolf (nominated by Eugene Koonin),
        David Krakauer, and Eugene Shakhnovich.}
\end{abstract}

\ifthenelse{\boolean{publ}}{\begin{multicols}{2}}{}
\section*{Open peer review}

This article was reviewed by  Eugene Koonin, Yuri Wolf (nominated by Eugene Koonin),
David Krakauer, and Eugene Shakhnovich.
%
\section*{Background}

The recent availability of complete genomic sequences
of a diverse group of living organisms allows one to
quantify basic mechanisms of molecular evolution on an
unprecedented scale. The part of the genome consisting
of all protein-coding genes (the full repertoire of
its proteome) is at the heart of all processes taking
place in a given organism. Therefore, it is very
important to understand and quantify the rates and
other parameters of basic evolutionary processes
shaping thus defined proteome. The most important of
those processes are:
\begin{itemize}
\item Gene duplications that give rise to
new protein-coding regions in the genome.
The two initially identical proteins encoded by a pair
of duplicated genes subsequently diverge from each
other in both their sequences and functions.

\item Gene deletions in which genes that are no longer
required for the functioning of the organism are
either explicitly deleted from the genome or stop
being transcribed and become pseudogenes whose
homology to the existing functional genes is rapidly
obliterated by mutations.
\item Changes in amino-acid sequences of proteins
encoded by already existing genes. This includes a
broad spectrum of processes including point
substitutions, insertions and deletions (indels), and
transfers of whole domains either from other genes in
the same genome or even from genomes of other species.
\end{itemize}

The BLAST (blastp) algorithm \cite{Altschul90}
allows one to quickly obtain the list of pairs of paralogous
proteins encoded in a given genome whose amino-acid sequences
haven't diverged beyond recognition.
The set of their percentage identities (PIDs) is a dynamic entity that changes
due to gene duplications, deletions, and local changes
of sequences.  Duplication events
constantly create new pairs of paralogous proteins with PID=100\%
, while subsequent substitutions, insertions and deletions result
in their PID drifting down towards lower values. A paralogous pair
disappears from this dataset if one of its constituent genes is
deleted from the genome, becomes a pseudogene, or when the PID of
the pair becomes too low for it to pass the E-value cutoff of the
algorithm. Thus the PID histogram contains a valuable if indirect
information about past duplications, deletions, and sequence
divergence events that took place in the genome.  In what follows
we propose a mathematical framework allowing one to extract some
of this information and quantify the average rates and other
parameters of the basic evolutionary processes shaping
protein-coding contents of a genome.

The list of all paralogous pairs
generated by the all-to-all alignment of protein sequences encoded
in a given genome is generally much larger than the list of
pairs of sibling proteins created by individual
duplication events.
For example, a family consisting of $F$ paralogous proteins
contributes up to $F(F-1)/2$ pairs to the
all-to-all BLAST output, while not more than $F-1$
of these pairs connect the actual siblings to each other.
The identification of the most likely candidates
for these ``true'' duplicates is in general a
rather complicated task which involves reconstructing the
actual phylogenetic tree for every family in a
genome. This goes beyond the scope of this study,
where we employ a much simpler (yet less precise)
Minimum Spanning Tree algorithm
to extract a putative non-redundant subset of true
duplicated (sibling) pairs.

The idea of quantifying evolutionary parameters using the
histogram of some measure of sequence similarity of duplicated
genes in itself is not new. It was already discussed
by Gillespie (see \cite{Gillespie_book} and references therein)
and later applied \cite{Lynch00} to
measure the deletion rate of recent duplicates.
There are two important differences between our methods
and those of the Ref. \cite{Lynch00}:
\begin{itemize}
\item
We use relatively slow changes in amino-acid sequences of proteins
as opposed to much faster silent
substitutions of nucleotides used in the Ref. \cite{Lynch00}.
This allows to dramatically extend the range of evolutionary times amenable to
this type of analysis.
\item
In addition to PID distributions in the
non-redundant set of true duplicated pairs
used in the Ref. \cite{Lynch00} we also study
that in the highly
redundant set of all paralogous pairs detected by BLAST.
It turned out that both these distributions
contain important and often
complimentary information about the quantitative dynamics of
the underlying evolutionary process. The shape
of the latter (all-to-all) histogram is to a
first approximation independent of duplication and deletion
rates and thus it allows us to concentrate on fine
properties of amino-acid substitution.
\end{itemize}

The central results of our analysis are:
\begin{itemize}
\item The middle part of the PID histogram of
all paralogous pairs detected by BLAST
is well described by a powerlaw
functional form with a nearly universal value of the exponent
$\gamma \simeq -4$ observed in a broad variety of genomes.
Our mathematical model relates this exponent to
parameters of intra-protein variability of
sequence divergence rates.
\item The upper part of the PID histogram corresponding to
recently duplicated pairs (PID$>$90\%) deviates from
this powerlaw form. It is exactly this subset of paralogous pairs
that was extensively analyzed in Ref. \cite{Lynch00}. This feature is
consistent with the picture
of frequent deletion of recent duplicates proposed in
Ref. \cite{Lynch00}.
\item The analysis of various features of
the PID histogram of all paralogous pairs and that of
a subset consisting of true duplicated (sibling) pairs allows us to
quantify both the long-term average duplication and deletion rates
in a given genome as well as
a dramatic increase in those rates for recently
duplicated genes.
\item
Abnormally flat PID histograms
observed for yeast and human are consistent with
lineages leading to these organisms undergoing one or more Whole-Genome
Duplications (WGD). This interpretation is corroborated by
the genome of {\it Paramecium tetraurelia} where
the PID$^{-4}$ profile of the sequence identity histogram is
gradually restored by the successive removal of paralogs
generated in its four known WGD events.
\item
Applying the same methods to large
individual families of paralogous proteins
allows one to study the
variability of evolutionary parameters within
a given genome. It is shown that larger or slower evolving
families are characterized by higher inter-protein
variability of amino-acid substitution rates.
\end{itemize}
%
\newpage
\section*{Results}
\subsection*{Distribution of sequence identities of all
paralogous pairs in a genome}

We studied the distribution of Percent Identity (PID)
of amino acid sequences of all pairs of paralogous
proteins in complete genomes of bacteria
{\it H. pylori, E. coli}, a single-celled eukaryote
{\it S. cerevisiae}, and multi-cellular eukaryotes
{\it C. elegans, D. melanogaster}, and {\it H. sapiens}.
Every protein sequence contained in a given genome was
attempted to be aligned with all other sequences in
the same genome using the blastp algorithm
\cite{Altschul90}.
To avoid including pairs of multidomain proteins
homologous over only one of their domains we have
filtered the data by only keeping the pairs in which
the length of the aligned region constitutes at least
80\% of the length of the longer protein. Infrequent
spurious alignments between different splicing
variants of the same gene or between proteins listed
in the database under several different names were
dropped from our final dataset. The exact details of
our procedure are described in the Methods section.
Fig. 1A shows the histogram $N_{a}(p)$ of amino-acid
sequence identities (PIDs) $p$ of {\it all pairs} of paralogous
proteins encoded in different genomes.  The $p$-dependence of
these histograms has three distinct regions I,II,III.
\begin{itemize}
\item
{\it Region I}: There is a sharp and significant upturn in the PID histogram
above roughly 90-95\% compared to what one expects from extrapolating $N_a(p)$ from lower
values of $p$. Apparently the constants (or possibly even
mechanisms) of the dynamical process shaping $N_a(p)$ are different in
this region.
\item
{\it Region II}: This region covers the widest interval of PIDs 30\%
$<p<$ 90\%.  $N_{a}(p)$ in this region can be approximated by a
power-law form of $p^{-\gamma}$ with $\gamma \approx 4$ (shown as a
dashed line in Fig. 1A.)
The best fits to the power-law form in the Region II
are listed in Table 2 and (with the exception of yeast
and human ) they fall in the $3-5$ range. The
near-universality of the shape of the PID histogram is
perhaps best illustrated by an approximate collapse of
PID histograms in different genomes when they are
normalized by the total number $N_{\mathrm{genes}}(N_{\mathrm{genes}}-1)/2$,
of all (both paralogous and non-paralogous) gene
pairs in the genome (Fig. 1B).
\item
{\it Region III}: In this region $p<25-30$\% the histogram $N_{a}(p)$
starts to deviate down from the $p^{-\gamma}$ powerlaw behavior. This
decline is an artifact of the inability of sequence-based algorithms
such as BLAST to detect some of the bona fide
paralogous pairs with low sequence
identity.  This explanation is corroborated by the observation that
the exact position of the downturn of $N_a(p)$ in the region III is
determined by the E-value cutoff (see Fig. 1S in the supplementary
materials).
\end{itemize}

\subsection*{Birth-and-death model of the proteome evolution}

In an attempt to interpret the empirical features of the PID
distribution described above we propose a simple stochastic birth and death
model of the proteome evolution.  It consists of a sequence of random gene
duplications, deletions, and changes in amino-acid sequences of proteins they
encode. Several versions of such
models were previously studied \cite{Huynen98,Gerstein01,Koonin02,Shakhnovich02}
most recently in the context of powerlaw distribution of family sizes. Our
model extends these previous attempts by concentrating
on evolution of sequence identities as opposed to just
the number of  proteins in different families.

Amino acid substitutions, insertions and deletions cause the sequence
identity of any given pair of paralogous proteins to decay with
time. Consider two paralogous proteins with PID=$p \times$100\% aligned against
each other. In the simplest possible case changes in their
sequences happen uniformly at all amino acid positions at a constant rate $\mu =
\mathrm{const}$. The effective ``substitution'' rate $\mu$ combines
the effects of actual substitutions and short indels.  The PID of this
paralogous pair changes according to the equation $dp/dt=-2\mu p$. The
factor two in the right hand side of this equation comes from the fact
that substitutions can happen in any of the two proteins involved, while the
factor $p$ - from the observation that only changes in parts of the two sequences
that remain identical at the time of the given change lead to a further decrease of the PID. This
equation results in an exponentially decaying PID: $p(t)\sim
\exp(-2\mu t)$. More generally the drift of PID could be described by
the equation $dp/dt=-v(p)$. When substitution rate varies for
different amino acids within the same protein the relationship between
$v(p)$ and $p$ would in general be non-linear. For our immediate
purposes we will leave it unspecified.  The negative drift of PIDs
generates a $p$-dependent flux of paralogous pairs down the PID axis
given by $v(p)N_a(p)$. The net flux into the PID bin of the width $\Delta
p$ centered around $p$ is given by $\Delta p \frac{\partial}{\partial p}[v(p)N_a(p)]$
(see Fig.2A)


Our model also involves random gene duplication and deletion events
(birth and death of new protein-coding genes)
which happen at rates $r_{\mathrm{dup}}$ and
$r_{\mathrm{del}}$ correspondingly.  In Fig.2B we
illustrate the details of how a gene duplication event creates new
pairs of paralogous proteins. When a gene A is duplicated to A' a new
pair of paralogs with PID=100\% is created (dotted line) and added to
the rightmost bin of the PID histogram. Furthermore the freshly
created gene A' inherits both paralogous partners (B and C) of the
gene A. The PIDs of these two newly created paralogous pairs A'-B
and A'-C (dashed lines) are also added to the respective bins in the
histogram.  Thus a duplication of any of the two paralogous genes with
PID=$p$ among other things results in the creation of a new pair of
paralogs with the same PID. This process increases $N_a(p)$ at a rate
$2r_{\mathrm{dup}} N_a(p)$. Similarly the deletion of any of the
two genes in this paralogous pair decreases $N_a(p)$ at the rate
$2r_{\mathrm{del}} N_a(p)$. The bin containing PID=100\%
($p=1$) has an
extra flux term $r_{\mathrm{dup}}N_{\mathrm{genes}}$
from PIDs of the freshly created pair of duplicated genes
(A-A' in our example). Here $N_{\mathrm{genes}}$ is the total
number of protein-coding genes in the genome.
Adding up contributions of the three main processes (substitutions,
duplications and deletions) one gets
%
\begin{eqnarray}
\frac{\partial N_a(p,t)}{\partial t}&=&\frac{\partial}{\partial
p}[v(p)N_a(p,t)]\nonumber
\\&+&2r_{\mathrm{dup}}N_a(p,t)-2r_{\mathrm{del}}N_a(p,t)+
r_{\mathrm{dup}}N_{\mathrm{genes}}\delta(p-1).
\nonumber
\end{eqnarray}
In our model the total number of genes $N_{\mathrm{genes}}$ in the
genome exponentially grows (or decays) according to
$dN_{\mathrm{genes}}/dt=(r_{\mathrm{dup}}-r_{\mathrm{del}})N_{\mathrm{genes}}$.
When the genome size of an organism remains (approximately) constant with
time one can find the stationary asymptotic solution of the previous equation. In
this case one must have $r_{\mathrm{dup}}=r_{\mathrm{del}}$ so
that the second term in the right hand side
is equal to zero.

In the case of an exponentially growing or shrinking
genome the stationary solution for
$N_a(p,t)$ does not exist.
However, it exists for the histogram normalized by the
total number of protein pairs:
%
It is easy to show that since
$\partial n_a(p,t)/\partial t=(\partial N_a(p,t)/\partial
t)/[N_{\mathrm{genes}}(N_{\mathrm{genes}}-1)/2]-
2(r_{\mathrm{dup}}-r_{\mathrm{del}})N_a(p,t)/[N_{\mathrm{genes}}(N_{\mathrm{genes}}-1)/2]$
the equation for normalized PID histogram $n_a(p,t)$ acquires an extra negative
term
$-2(r_{\mathrm{dup}}-r_{\mathrm{del}})n_a(p,t)$.
This term exactly cancels the duplication and deletion terms in
the equation for $N_a(p,t)$ and considerably
simplifies the equation for the normalized histogram :
\begin{eqnarray}
\frac{\partial n_a(p,t)}{\partial t}&=&\frac{\partial}{\partial
p}[v(p)n_a(p,t)]+\frac{2r_{\mathrm{dup}}}{N_{\mathrm{genes}}-1}\delta(p-1).
\label{eq:1}
\end{eqnarray}
In the steady state solution one has $\partial n_a(p,t)/ \partial t=0=\partial /\partial p
[v(p)n_a(p)]$ or
\begin{equation}
n_a(p) \sim N_a(p)\sim 1/v(p).
\label{eq:N_a}
\end{equation}
%

The conjecture that the normalized PID
histogram $n_a(p)=N_a(p)/[N_{\mathrm{genes}}(N_{\mathrm{genes}}-1)/2]$
indeed is nearly stationary during the course of
evolution is corroborated
by the fact that all six $n_a(p)$ curves in
various genomes used in our study approximately
lie on top of each other in Fig. 1B (compared to
unnormalized $N_a(p)$ shown in Fig.
1A).

Comparing the Eq.~\ref{eq:N_a} with the empirical form of $N_a(p) \sim
1/p^4$ in the region II of Fig. 1
one concludes that the drift velocity in real genomes must obey
$v(p)\sim p^4$. Such a non-linear dependence of $v(p)$ could be
explained by the variability of the effective substitution rate within
proteins (intra-protein variability).  Assuming the intra-protein
variability of substitution rates $\mu$ described by a PDF $\rho(\mu)$
one gets the following expression for $p(t)$ and $v(t)$:
\begin{eqnarray}
\label{eq:p(t)}
p(t)&=&\int_0^{\infty} \rho(\mu) e^{-2\mu t} d\mu \quad ; \\
v(t)&=&-\frac{dp(t)}{dt} = \int_0^{\infty} 2 \mu \rho(\mu) e^{-2\mu t} d\mu
\quad .
\label{eq:v(t)}
\end{eqnarray}
Eq.~\ref{eq:p(t)} is a generalization of the previously discussed
exponential decay of $p(t)$ derived for a constant substitution rate
$\mu$. It simply weighs these exponentials by $\rho (\mu)$ - their
likelihood of occurrence. For any given $\rho (\mu)$ one could exclude
time from Eqs.~\ref{eq:p(t)} and \ref{eq:v(t)} and express $v$ as a
function of $p$. Such $v(p)$ dependence could then be directly
compared with the empirically derived formula. In the absence of an
analytical expression relating $v(p)$ to $\rho(\mu)$ one is limited to
use a trial-and-error method.  We start with Gamma-distributed
$\rho(\mu)
\sim \mu^{\alpha-1}\exp (-\mu/\mu_0)$ which has been
predominantly used in the literature
(\cite{Zhang98,Yang93,Grishin00}).
Inserting thus defined $\rho(\mu)$ into Eqs.~\ref{eq:p(t)} and
\ref{eq:v(t)} one gets $p(t)=(2\mu_0)^{-\alpha}/(t+(2\mu_0)^{-1})^\alpha$ and
$v(t)=\alpha(2\mu_0)^{-\alpha}/(t+(2\mu_0)^{-1})^{\alpha+1}$ which leads to $v(p)\sim
p^{(\alpha+1)/\alpha}$ and thus to $N_a(p)\sim p^{-(\alpha+1)/\alpha}$.

\subsection*{Robustness of the functional form of
$N_a(p)$ with respect to assumptions used in the
model.}
The birth-and-death model described above is based on
a simplified picture of genome evolution.
In particular it implicitly
assumes:
\begin{itemize}
\item
The neutrality of individual gene duplication and deletion events resulting in
identical rates of these two processes in all paralogous families in the genome.
\item
Identical average amino-acid substitution rates $\mu_0$
in all individual proteins.
\end{itemize}
Both of these assumptions are known to be not, strictly speaking, true.
Sequences of some ``important'' proteins (e.g. constituents
of the ribosome) are known to evolve very slowly.
Also, the families containing essential (lethal knockout)
genes were recently shown \cite{Boris_Koonin_07}
to be characterized by higher average duplication and deletion
rates than those lacking such genes.

However, the validity of our main
results goes well beyond the validity of the
approximations that went into our birth-and-death
model. The advantage of using the histogram of sequence
identities generated by the all-to-all alignment ($N_a(p)$)
lies in its remarkable universality and robustness.
When the  Eq. \ref{eq:N_a} is applied to individual families one can see
that family-to-family variation of (and
correlations between) the duplication rate $r_{\mathrm{dup}}$, the deletion rate
$r_{\mathrm{del}}$, and the average substitution rate
$\mu_0$ affect only the prefactor
in the powerlaw form of $N_a(p)$.
Thus the exponent
$\gamma=1+1/\alpha$ describing this powerlaw
is very robust with respect to assumptions of the model
and depends only to the exponent $\alpha$ quantifying the
{\it intra-protein} variability of
amino-acid substitution rates.

The exact mechanisms behind this apparent universality of
$\alpha$
are not entirely clear. Chances are that it is
dictated more by the protein physics
rather than by organism-specific evolutionary mechanisms.
A possible path towards derivation of the exponent
$\alpha$ from purely biophysical principles
starts with the results of
Ref. \cite{Dokholyan_Shakhnovich_2001}, which models
the effects of (correlated) multiple amino-acid substitutions
on stability of the native state of a protein. However, such
analysis goes beyond the scope of the present
work and will be reserved for a future study.

%
%
\subsection*{Distribution of sequence identities of true duplicated pairs}

A highly redundant dataset consisting of all paralogous pairs present
in the genome enabled us to quantify the variability of intra-protein
substitution rates. Another set of important parameters describing proteome
evolution are average deletion/inactivation and duplication rates
$r_{\mathrm{del}}$ and $r_{\mathrm{dup}}$. As will be shown in the following, the
reduced non-redundant dataset consisting only of protein pairs
directly produced in duplication events allows us to
estimate these rates.

To better understand the difference between those two datasets we
illustrate it with a simple example. The family of four evolutionary
related proteins A,B,C,D contributes six paralogous pairs to
$N_a(p)$. This family was actually created by three subsequent
duplication events: first A duplicated to give rise to B, then B
duplicated to C and finally C duplicated to D. Thus only three out of
total six paralogous pairs are directly produced in gene duplication
events. The actual
number of duplicated pairs could be even smaller if some intermediate
genes were deleted in the course of the evolution.  In general a
family consisting of $F$ proteins contributes at or around $F(F-1)/2$
paralogous pairs to $N_a(p)$, but only $F-1$ duplicated pairs to
$N_d(p)$.

Nothing in the BLAST output for a given paralogous pair contains any
information if it should or should not be included in the $N_d(p)$.
However, using the set of all sequence identities of proteins for a
given family one could tentatively reconstruct the course of
duplication events that led to the appearance of this family.
Generally speaking, this is a rather complicated task involving
reconstructing the actual phylogenetic tree for every family in a
genome. In this study we use a much simpler
alternative based on the Minimum Spanning Tree (MST) algorithm (see
Methods for more details). For each protein
family this algorithm generates a tentative set of duplication
events in its past history. Numbers of pairs included in $N_a(p)$ and
$N_d(p)$ distributions in different organisms are listed
in the Table 1.

The dynamics of the distribution of duplicated pairs $N_d(p,t)$ is
described by simply excluding the duplication term $2
r_{\mathrm{dup}} N(p, t)$ from the equation for
$N_a(p, t)$.  Indeed, this
term is caused by PIDs of non-duplicated paralogs
A'-B and A'-C (dashed lines
in Fig. 2
(B)) generated when a gene A was duplicated. However, only the actual
duplicated pair A-A' with initial PID of 100\% (dotted line in Fig. 2
(B)) is included in the distribution of duplicated pairs
$N_d(p)$. Thus the dynamics of $N_d$ is described by
%
$$
\frac{\partial N_d(p,t)}{\partial t}=\frac{\partial}{\partial p}
[v(p)N_d(p,t)]-2 r_{\mathrm{del}}N_d(p, t)+
r_{\mathrm{dup}}N_{\mathrm{genes}}\delta(p-1). \quad.
$$
%
Once again the stationary solution exists for the
normalized distribution. However, in this case the
correct normalization factor is given by $N_{\mathrm{genes}}$
and not $N_{\mathrm{genes}}(N_{\mathrm{genes}}-1)/2$ as for $N_a(p)$.
Indeed, every duplication event increasing the number of genes by one
adds just one duplicated pair to $N_d(p)$ but up to
$N_{\mathrm{genes}}$ pairs to $N_a(p)$.
Thus the normalized PID
histogram of duplicated pairs
$n_d(p,t)=N_d(p,t)/N_{\mathrm{genes}}$ evolves
according to
\begin{equation}
\frac{\partial n_d(p,t)}{\partial t}=\frac{\partial}{\partial p}
[v(p)n_d(p,t)]-(r_{\mathrm{dup}}+r_{\mathrm{del}})n_d(p, t)
+r_{\mathrm{dup}}\delta(p-1) \quad.
\label{N_d}
\end{equation}
According to our empirical findings the average rate of sequence
divergence of paralogous proteins in most organisms is described
$v(p)=2\bar{\mu} p^{\gamma} $, where $\bar{\mu}$ is the substitution
rate averaged over all amino-acid positions in all proteins, and
$\gamma \approx 4$ is the exponent related to the intra-protein
variability of $\mu$.  The steady state of Eq. \ref{N_d}: $\partial
n_d/\partial t=0$ is satisfied by:
\begin{equation}
N_d(p) \sim n_d(p)=\frac{r_{\mathrm{dup}}}{2\bar{\mu}p^{\gamma}}
\exp\left(-\frac{r_{\mathrm{dup}}+r_{\mathrm{del}}}{2\bar{\mu}(\gamma-1)p^{\gamma-1}}\right).
\label{eq:Ndfinal}
\end{equation}
\subsection*{Numerical test of analytical predictions}

The analytical results derived above were confirmed by a numerical
simulation.  An artificial ``proteome'' used in our numerical model
consists of a fixed number of ``proteins'' of identical lengths. At
every timestep one takes the sequence of a randomly selected protein
and uses it to overwrite the sequence of another randomly selected
protein. This corresponds to a stationary genome case
when $r_{\mathrm{dup}}=r_{\mathrm{del}}$.
Each combined duplication/deletion event is followed by
random substitutions of several ``amino-acids'' (see Methods for details of
our simulation).  In the beginning of the simulation each amino-acid
position within every protein was randomly assigned a substitution
rate drawn from the Gamma-distribution with $\alpha=1/3$. The proteome
generated by this dynamical process is periodically analyzed in terms
of sequence identity of all pairs of its proteins. The resulting
distributions $N_a(p)$ (filled circles) and $N_d(p)$ (open circles)
are shown in Fig. 3.
They agree quite well with our
theoretical predictions: $N_a(p) \sim 1/p^4$ (solid line) and $N_d(p)
\sim 1/p^4\exp(-r_{\mathrm{del}}/[3\mu p^3])$ (dashed line).
The best fit to $N_d(p)/N_a(p)$ generated in our numerical
simulation with $A\exp(-r_{\mathrm{del}}/[3\mu p^3])$
gives $r_{\mathrm{del}}/\mu = 0.237$
in excellent agreement with the actual value of $r_{\mathrm{del}}/\mu = 0.25$
used in our simulation. This demonstrates that
evolutionary parameters can be successfully
reconstructed from the shapes of $N_d(p)$ and
$N_a(p)$. This is especially encouraging in case
of $N_d(p)$ because of the approximations
that went into identifying true duplicated (sibling) pairs by
the Minimum Spanning Tree algorithm.
\subsection*{Fitting evolutionary parameters of real proteomes:
the long-term deletion rate.}

To estimate the average of deletion and duplication
rates we performed a two-parameter fit to the $N_d(p)/N_a(p)$ ratio
with
$A\exp(-B/(\gamma-1)p^{\gamma-1})$
(see Eqs. \ref{eq:N_a},\ref{eq:Ndfinal})
in the $30\%<p<90\%$ interval (region II in Fig. 1 ).
Here $A$ and $B=(r_{\mathrm{dup}}+r_{\mathrm{del}})/(2\bar{\mu})$
are the two free fitting parameters. The exponent
$\gamma$ used in the fitting formula itself
was obtained from the best fit
to $N_a(p)$  in the same region with the power-law
form $p^{-\gamma}$ (see column 3 in Table 1). The
ratio $r_{\mathrm{del}}/\bar{\mu}$
was extracted from the best-fit value of $B$ and the
independently calculated
duplication rate ratio $r_{\mathrm{dup}}/\bar{\mu}$
(see subsection below). It is listed in the
sixth column of the Table 2.
\subsection*{Fitting evolutionary parameters of real proteomes:
the short-term deletion rate of recent duplicates.}

A very pronounced and reproducible feature in all
organism-wide histograms   is an abrupt drop as   is
lowered from 100\% down to about 90-95\% (region I in
Fig.1.) The drop is as large as 30-fold in prokaryotes
and is around 3-to-10 fold in eukaryotes. It is
subsequently followed by a   increase of   in the
region II which at low  25\% (region III) turns down
again only due to limitations of our ability to detect
evolutionary related sequences. There exists several
possible explanations for this initial drop in the
region I :
\begin{itemize}
\item The gene conversion process.
In a gene conversion process a part or
the whole sequence of one of the paralogous genes is
used as a template to modify the sequence of another.
It happens with a reasonable frequency only if those
two genes are sufficiently close to each other
in their sequences
so that DNA repair mechanisms
might mistakenly assume that one of them is the corrupted version of
the other. If gene conversion events are sufficiently
common, the initial separation of a pair of freshly
duplicated genes may take a long time, as one of them
would be getting constantly converted back to the
other. This would result in an abnormally small drift
velocity  $v(p)$ for $p$ close to 100\% and hence to
an abnormally high $N_a(p) \sim 1/v(p)$.
\item
Another, more plausible
explanation is that freshly duplicated genes are
characterized by a much higher deletion rate
$r_{\mathrm{del}}^* \gg r_{\mathrm{del}}$
\cite{Lynch00}.
Functional roles of such genes have not had enough
time to diverge from each other making each of them
more disposable than an average gene in the genome.
Indeed, for {\it S. cerevisiae} and {\it C. elegans} it was
empirically demonstrated \cite{Gu03,Maslov04}
that the deletion or inactivation of genes with a
highly similar paralogous partner in the genome is up
to 4 times more likely to have no consequences for the
survival of the organism than the
deletion/inactivation of genes lacking such a partner.
\end{itemize}
The $N_a$ dynamics in the region I ($p \simeq
100$\%) is then described by
$$
\frac{\partial N_a(p, t)}{\partial t}=\frac{\partial}{\partial p}[2\bar{\mu} N_a(p, t)]
+(2r_{\mathrm{dup}}- 2r_{\mathrm{del}}^* )
N_a(p,t)+
r_{\mathrm{dup}}N_{\mathrm{genes}}\delta(p-1) \qquad .
$$
while the normalized distribution
$n_a=N_a/[N_{\mathrm{genes}}(N_{\mathrm{genes}}-1)/2]$ obeys
\begin{equation}
\frac{\partial n_a(p, t)}{\partial t}=\frac{\partial}{\partial p}[2\bar{\mu} n_a(p, t)]
+(2r_{\mathrm{del}}- 2r_{\mathrm{del}}^* ) n_a(p,t)+
\frac{2r_{\mathrm{dup}}}{N_{\mathrm{genes}}-1}\delta(p-1) \qquad  .
\label{eq:regionI}
\end{equation}
%
Here $2\bar{\mu}=v(100\%)$ is the average substitution rate in freshly
duplicated pairs and $r_{\mathrm{del}}^*$
is the deletion rate inside region I.
The equation has an exponentially decaying
stationary solution which for
$r_{\mathrm{del}}^* \gg r_{\mathrm{del}}$ is
simply given by $n_a(p) \sim \exp
(r_{\mathrm{del}}^*
p/\bar{\mu})$ This functional form is consistent with the empirical
data for $p$ just below 100\%
and the best fits to $r_{\mathrm{del}}^*/\bar{\mu}$
are listed in the seventh column of the Table 2.
Ref. \cite{Lynch00}
analyzed the distribution of silent substitution numbers per silent site  $K_s$
between pairs of recently duplicated genes.
Under the same ``drift and deletion'' hypothesis used to
derive the Eq. \ref{eq:regionI} such $K_s$-distribution $N_d(K_s)$
should also have an exponential decaying form $N_d(K_s) \sim \exp
(-r_{\mathrm{del}}^* K_s/\bar{\mu}_s)$, where $\bar{\mu}_s$ is the
average drift velocity of $K_s$ immediately following the duplication
event. Fits to this exponential functional form performed in Ref. \cite{Lynch00}
resulted in $r_{\mathrm{del}}^*/\bar{\mu}_s \sim
7-24$. Our estimates $r_{\mathrm{del}}^*/\bar{\mu} \sim 20-70$
are consistent with those of (\cite{Lynch00}
) provided that the
$\bar{\mu}/\bar{\mu}_s$ ratio is in $0.1-1$ interval.
\subsection*{Fitting evolutionary parameters of real proteomes:
long- and short-term duplication rates.}

The number of
paralogous pairs with PID$\simeq$100\% also contains
information about the raw duplication rate $r_{\mathrm{dup}}^*$ in the
genome. This rate is subsequently trimmed down to its
long-term stationary value $r_{\mathrm{dup}}$  by the removal of a large fraction
of freshly created pairs as described in the previous subsection.
New pairs with
PID=100\% are created at a rate  $r_{\mathrm{dup}}^*N_{\mathrm{genes}}$,
 while they leave the
bin containing PID=100\% at a rate
$2\bar{\mu}N_a(100\%)/\Delta p$ . Here $\Delta p$ is the
width of the bin and $N_a(100\%)$ is the number of pairs in this
last bin. The width of the bin is assumed to be small
enough so that the removal of genes from the bin due
to deletion is negligible in comparison to that due to
the drift in their sequences. Thus $r_{\mathrm{dup}}^*/\bar{\mu}=
2N_a(100\%)/(N_{\mathrm{genes}}\Delta p)$ . The average
duplication rates calculated this way are presented in
the fourth column of Table 2 . They are compatible
with  $r_{\mathrm{dup}}^*/\bar{\mu}_s$
calculated in \cite{Gu02}, where the same idea was
applied to $N_d(K_s)$.

The rate  $r_{\mathrm{dup}}^*$ includes the creation
of some extra duplicated pairs which are then quickly
(on an evolutionary timescale) eliminated from the
genome during a ``trial period'' while their PID$>$90\%. We have
already demonstrated that such a deletion happens at a
very high rate  $r_{\mathrm{del}}^*$
and thus has to be treated separately
from the background deletion rate $r_{\mathrm{del}}$.
The duplication
rapidly followed by a deletion does not change the
overall distribution of paralogous pairs. Therefore,
the long-term average
duplication rate $r_{\mathrm{dup}}$
used in Eqs. 2, 6 is in fact
considerably lower than the raw duplication rate
$r_{\mathrm{dup}}^*$. An approximate
way to calculate it is to use power-law fits
to $N_a(p)$ in the region
II to extrapolate it up to 100\%. Such extrapolated
value $N_a^{\mathrm{ext}}(100\%)$ could then be used
to calculate the long-term average duplication rate
as $r_{\mathrm{dup}}/\bar{\mu}=
2N_a^{\mathrm{ext}}(100\%)/(N_{\mathrm{genes}}\Delta p)$.
(see the fifth column of
Table 2).
%
\section*{Discussion and Conclusions}
\subsection*{An estimate of the number of superfamilies in different
genomes}
Any sequence-based method is bound to miss similarities
between some of the distant paralogs.
The situation could be somewhat improved \cite{Shakhnovich02}
if one compares proteins' three-dimensional structures which
are conserved over longer evolutionary times.
The addition of  previously undetected paralogous pairs
results in some of the sequence-based families merging together to
form larger superfamilies.
Our empirical observations
allow us to estimate the number of such
superfamilies contained in a given genome. Indeed, the fraction of
paralogous pairs among all gene pairs in a genome
consisting of $N_F$ mutually unrelated superfamilies
is given by $1/N_F$. A rough estimate
of $N_F$ is provided
by extrapolation of the $p^{-4}$ powerlaw into the Region III
down to some cutoff $\mathrm{PID}_{\mathrm{min}}$:
\begin{equation}
N_F=\frac{N_{\mathrm{genes}}(N_{\mathrm{genes}}-1)}{2}
\Big /
\int_{\mathrm{PID}_{\mathrm{min}}}^{1} Ap^{-4}dp
\qquad .
\label{eq:N_F}
\end{equation}
Here A is the best fit to $N_a(p)$ with the $p^{-4}$
in the region II.
Remarkably, the results of such calculation are roughly
genome independent.
Using the lowest
theoretical limit PID$_{\mathrm{min}}$=5\%
(the sequence identity of two
unrelated sequences composed of 20 amino-acids)
results in the effective number of superfamilies
$N_F$ ranging between 4.7 in {\it C. elegans}
and 9.9 in {\it D. melanogaster}.
A more realistic limit
PID$_{\mathrm{min}}$=8\% \cite{Shakhnovich02},
which takes into account the non-uniform frequency among 20
amino-acids, somewhat increases the number of
superfamilies to 36,  28,  31,  19,  40, and  35 for
{\it H. pylori, E. coli, S. cerevisiae, C.
elegans, D. melanogaster}, and {\it H. sapiens}
correspondingly.
These numbers are still respectably small compared to
$N_F \simeq 1000$ one gets by using the cutoff
PID$_{\mathrm{min}}$=25\%
imposed by the inadequacy of sequence-based methods
to detect similarity of remote paralogs.

\subsection*{The exponent $\alpha$ in large individual
families}


The Gamma-distribution
$\sim \mu^{\alpha-1}\exp (-2 \mu/\mu_0)$
was traditionally used
to model and fit the distribution of substitution rates in
individual families of proteins
(this tradition goes back to
\cite{Uzzell71}).
Our approach extends this approach to a proteome-wide
scale and demonstrates that beyond its role as a
{\it ad hoc} fitting function the
Gamma-distribution indeed provides an excellent
quantitative description of variability of intra-protein
substitution rates.

The genome-wide value of the exponent $\alpha \simeq 0.33$
obtained in our analysis is consistent with
its previous estimates in large protein families.
For example, the best fits with the Gamma-distribution
performed in Refs. \cite{Yang93,Grishin00}
resulted in the exponent $\alpha$
in the $0.2-0.4$ range.
The authors of Ref. \cite{Zhang98}
quantified the variability of substitution rates using a
large set of orthologous proteins in different
genomes (this should be contrasted with paralogous proteins
used in our analysis).
The fits with Gamma-distribution resulted in
a broad range of exponents $\alpha$ for individual proteins.
Still the distribution of exponents $\alpha$ peaks around 0.25
(solid histograms in the Fig.4 of Ref. \cite{Zhang98}) .

In principle, our methods could be also applied to
large individual families of paralogous proteins.
However, only a few of the largest families in
any genome contain sufficient number of
paralogous pairs (up to $F(F-1)/2$ in a family of size
$F$) to have a meaningful individual $N_a(p)$  histogram.
Fig. 5 shows the results of such analysis in
{\it C. elegans} with individual curves corresponding to
$N_a(p)$ in the 5 largest families of paralogous proteins.
Without exception all these histograms are well fitted by
$Cp^{-\gamma}$ with $\gamma_{LF} \simeq 5-6$. This
corresponds to the exponent $\alpha \simeq 0.17-0.25$, which is a somewhat
lower than the exponent
$\alpha=0.33$ that we observed for genomes as a whole.

Smaller individual families do not hold sufficient
statistical power to analyze the shape of $N_a(p)$. One
approach would be to group them together by some
shared characteristic (e.g by their size, or by whether or
not they contain an essential gene as described in Ref.
\cite{Boris_Koonin_07}). However, the $N_a(p)$
histogram in such a group would depend on additional
parameters such as the rate of creation and removal
of families of a given type and thus will not be amenable
to our type of analysis.
For example the collection of families binned by their size
would have additional birth-and-death events due to whole families
entering or leaving the selected bin.
The rates of these processes would have a non-trivial
dependence on the age of a family and thus cannot be
easily incorporated into our mathematical framework.

It is important to emphasize once again that
the exponent $\alpha$ quantifies only
the {\it intra-protein} variability of substitution rates
at different amino-acid positions within individual proteins.
Such variability should not be confused with
a much larger protein-to-protein
variability of average substitutions
rates. Indeed, sequences of different proteins encoded in the same genome are
known to evolve at vastly different rates (see \cite{Zhang98,Grishin00} and
references therein). Some sequences, such as e.g.
those of ribosomal proteins, remain virtually unchanged
over billions of years of evolution, while others
change at a much faster pace. In fact, the very
importance of a protein is sometimes quantified by its
average rate of evolution
as more essential proteins involved in core cellular processes
tend to evolve at slower than average rates.

\subsection*{Genome size dependence and other properties
of long- and short-term duplication and deletion rates}

Our data indicate that the long-term
duplication rate $r_{\mathrm{dup}}$
is of the same order of magnitude as the
long-term deletion rate $r_{\mathrm{del}}$
(see columns 5 and 6 in the Table 2).
This is to be expected since any large discrepancy in
these rates would generate much greater differences in
genome sizes than actually observed in these model organisms.
However, as was proposed by \cite{Lynch00},
both of these rates are considerably
smaller than their short-term (``raw'') counterparts
$r_{\mathrm{dup}}^*$ and $r_{\mathrm{del}}^*$ that include
recently duplicated proteins.

Our results for the fruit fly {\it D. melanogaster}
are consistent with an earlier observation \cite{Gu02}
of an abnormally low average duplication rate in
this organism.
According to our data $r_{\mathrm{dup}}^*/\bar{\mu}$
is about nine times lower than that in the genome of
{\it C. elegans}. The long-term stationary duplication
rate $r_{\mathrm{dup}}/\bar{\mu}$ in the fly is also the lowest in all eukaryotic
genomes used in this study but is only three times
lower than that in the worm.

Intriguingly, $r_{\mathrm{del}}/\bar{\mu}$,
$r_{\mathrm{dup}}/\bar{\mu}$, and
$r_{\mathrm{dup}}^*/\bar{\mu}$ ratios are all
positively correlated with the complexity of the
organism quantified by the total number of genes in
its genome (see correspondingly filled circles, open diamonds, and
open squares in Fig. 4).
This means
that either the per-gene duplication rate in
more complex organisms is consistently higher than in
their simpler counterparts or that their average
amino-acid substitution rate is lower. It is likely
that both above trends operate simultaneously. One possible
explanation for the latter trend is that
the more sophisticated mechanisms of DNA copying and
repair of higher organisms lead to lower average
amino-acid substitution rates.

On the other hand,
we find that the deletion rate of recent
duplicates, $r_{\mathrm{del}}^*/\bar{\mu}$,
(filled triangles in Fig. 4) is
negatively correlated with the number of genes in the
genome. This result is in agreement with
Ref. \cite{Lynch03} where this trend was attributed to the decrease
in effective population size in more complex
organisms.

\subsection*{The effects of whole genome duplications
on the histogram of sequence identities}
Two of the organisms used in our
study ({\it S. cerevisiae} and {\it H. sapiens}) are
characterized by a dramatically lower value of the power-law exponent
$\gamma$ (1.8 for yeast and 2.4 for human)
and the overall poor quality of the power law fit to $N_a(p)$.
One plausible explanation of this anomaly
is in terms of Whole Genome
Duplications (WGD) in lineages leading to these
genomes.
It is well established  \cite{wolfe1997mea} that
baker's yeast underwent a WGD event, which
most likely occurred about 100 Myrs
ago.
While the subject remains controversial,
it is also  commonly believed that the vertebrate lineage
leading to {\it H. sapiens} (among many other vertebrate genomes)
also underwent one or
several large-scale duplication events
\cite{lundin1993evg,Wolfe2002}.
%
In the immediate aftermath of a WGD event the PID
distribution changes as follows: $N_a(p) \to 4N_a(p)$
for p$<$100\%, while $N_a(100\%) \to
4N_a(100\%)+N_{\mathrm{genes}}$. Indeed, every
ancestral paralogous pair A-B would give rise to 3 new
pairs with the same PID: A-B', A'-B, and A'-B'. At the
same time the bin containing the PID=100\% would in
addition get $N_{\mathrm{genes}}$ (or fewer for a
large segmental duplication) of freshly created
duplicated pairs of the type A-A' and B-B'. The
subsequent spread of this sharp peak at PID=100\%
towards lower values of PID accompanied by a rapid
deletion of redundant copies of duplicated genes
would result in an effective flattening of the $N_a(p)$ histogram in its
upper range and thus in lower effective value of the
exponent $\gamma$.

To further test this hypothesis we analyzed the
recently sequenced genome \cite{ciliate_WGD} of
a ciliate {\it Paramecium tetraurelia}.
This organism underwent as many as four separately
identifiable WGD events \cite{ciliate_WGD}.
We used our standard methods to construct the  PID histogram
$N_a(p)$ from the all-to-all alignment of its
nearly 40,000 genes. Due to the sheer size of this
proteome we employed the same conservative
$10^{-30}$ E-value cutoff we used for
{\it H. sapiens} and {\it C. elegans}.
Solid diamonds in the Fig. 6 correspond to
the full PID histogram in {\it Paramecium tetraurelia}
consisting of all 103,828 paralogous pairs
detected by our methods.
Authors of Ref. \cite{ciliate_WGD} identified the
lists of putative pairs of
duplicated genes generated in each of the four WGD
events in the lineage leading to this genome.
By dropping one randomly-selected gene from these WGD pairs we
generated the set of four progressively more narrow
PID histograms. These histograms are also shown in
Fig. 6: 41,890 pairs excluding the genes generated in the latest
WGD event (solid squares), 25,342 pairs excluding the genes
generated in the last two WGD events (solid circles),
22287 pairs excluding the genes generated in the latest
three WGD events (open triangles), and 21,417 pairs
excluding the genes generated in all four WGD events (red
stars). For comparison, the Fig. 6 also reproduces
the histogram of 31,078 pairs in the $N_a(p)$ of {\it H.
sapiens} (blue $\times$-es). One can see that progressive elimination
of pairs generated in WGD events gives rise to the $N_a(p)$ histogram
approaching the universal scaling form:
$N_a(p) \sim p^{-4}$ (the dashed line in Fig.6).
Furthermore,  the PID distribution of
gene pairs generated in each of the WGD events has a
shape that is qualitatively consistent with the predictions of our
birth-and-death model. In particular, the gene pairs from the latest round
of WGD did not have time to sufficiently
diverge. As a result, their PID-distribution (shown in black in the Fig. 3a of the
Ref. \cite{ciliate_WGD}) has a peak around 95\% sequence identity with a
half-maximum at 75\%.
The analysis of {\it Paramecium tetraurelia} genome
provides an additional strong support to our conjecture that
the unusually flat PID histograms in human and baker's yeast
are caused by  WGD events in lineages leading to
these two organims.

%
%
%

\section*{Methods}
\subsection*{The details of generating lists of paralogous proteins}

The proteomes of {\it H. pylori} strain 26695 and {\it E.
coli} strain K12-MG1655 were downloaded from the
Comprehensive Microbial Resource (CMR) \cite{tigr}
version 1.0. Sequences of S. cerevisiae proteins are
from the Saccharomyces Genome Database (SGD)
\cite{SGD} version number 20031001. The {\it D.
melanogaster}'s sequences are from the Berkeley
Drosophila Genome Project \cite{flybase}, release
3.1. {\it C. elegans} - Wormbase \cite{wormbase}, release
WS127.{\it H. sapiens} - the NCBI database,\cite{NCBIdatabase},
build 34.1. The initial
set of paralogous pairs for each of the organisms was
identified by an all-to-all alignment of sequences of its proteins to
each other using the BLASTP program \cite{Altschul90}.
For {\it H. pylori, E. coli, S. cerevisiae}, and {\it D.
melanogaster} genomes, the E-value threshold of
$10^{-10}$ was employed. This corresponds to p-values of the
order of $10^{-12}$ (for {\it H. pylori}) and lower. Due to
larger genome sizes of {\it C. elegans} and {\it H. sapiens} an even more
conservative E-value of $10^{-30}$ was used to reduce the
number of hits generated by the algorithm.

The ``raw'' datasets for worm, fly and human often contain
multiple overlapping protein sequences predicted by
different gene models of the same gene (including but
not limited to different splicing variants). To avoid
spurious hits we first mapped entries in raw datasets
to unique gene IDs. This was easy to accomplish in the
fly and worm datasets, where names of different gene
models differ from each other by the last letter.  In
human genome, this was done by mapping the gi numbers
of sequences in the raw dataset to unique GeneID
(LocusLink) identifiers from the Entrez Gene database
\cite{Entrezgene}.
Subsequently, if multiple BLAST hits were connecting
the same pair of gene IDs we kept the one with the
longest aligned region. This way we were guaranteed
that one and only one pair of splicing (or gene model)
variants per pair of gene IDs would contribute to the
PID histogram.

In all genomes, only pairs in which the
aligned region constituted at least 80\% of the length
of the longer protein were kept \cite{Gu02}. This
excludes contribution from pairs of multi-domain
proteins paralogous over only one of their domains.

Initially, the PID histogram in {\it S. cerevisiae} had two
very sharp peaks at 51\% and 70\%. A close inspection
revealed that these peaks are produced by evolutionary
related subfamilies of nearly identical transposable
elements. To correct for this obvious artifact in {\it S.
cerevisiae} we removed 108 proteins encoded by known
transposable elements listed in the Saccharomyces Genome Database
\cite{SGD} and their homologs.

The overall shape of the PID histogram in regions I and II
is not sensitive to the E-value cutoff chosen. In Fig.
1S we show that when the E-value cutoff in the fly
dataset was changed from a less conservative $10^{-10}$ to
a more conservative $10^{-30}$ value, the shape of the
histogram above ~40\% remained virtually unchanged.
Similarly, the results are nearly independent on the
type of the BLOSUM substitution matrix used (in the
end we opted for the BLOSUM45.) Finally, we verified
that our results are independent of the alignment
algorithm utilized to calculate PIDs. Indeed, in the
fly dataset we have recalculated PIDs for all
paralogous pairs detected by BLAST using much more
sophisticated Smith-Waterman algorithm \cite{SmithWaterman81}.
The resulting histogram (shown as blue stars in
Fig. 1S) is virtually indistinguishable from that
based on the blastp output.
\subsection*{Numerical model of the proteome evolution}

We numerically simulated a birth and death model mimicking the
evolution of a fixed-size proteome by duplication,
deletion and substitutions. We first randomly fill a
2,000$\times$100 matrix with integer numbers ranging from 1
and 20 (20 types of ``amino-acids''). This constitutes
the initial state of our artificial genome/proteome,
encoding 2,000 ``proteins'' of 100 ``amino-acids'' each.
Every amino-acid position in each of the proteins is
randomly assigned the substitution rate $\mu$  drawn from a
Gamma-distribution with $\alpha=1/3$. One evolutionary timestep
consists of:
\begin{enumerate}
\item
Duplicate a randomly selected gene in the genome and
use this duplicated copy to replace another randomly
selected gene (deletion). Thus in this model the
deletion rate is exactly equal to the duplication
rate.
\item
Randomly pick 400 amino-acid positions in the whole
genome and substitute amino-acids at those positions
to a randomly selected new value. The probability of a
particular amino-acid position to be picked is
proportional to its substitution rate $\mu$.
\end{enumerate}
This choice
of parameters in our model corresponds to
$r_{del}/\bar{\mu}=r_{del}^*/\bar{\mu}=r_{dup}/\bar{\mu}=r_{dup}^*/\bar{\mu}=0.25$.
Indeed, the average substitution rate per amino acid during
one timestep is given by $400/(100\times 2000)=1/500$. It is equal to $0.25$
of the per-gene per timestep duplication/deletion rate of $1/2000$. In this artificial
evolutionary process we have the advantage of keeping
track of all the duplicated pairs. Thus, after each
duplication event the list of all duplicated pairs is
updated and can be directly read off. After repeating
the above steps for 20,000 times the full genome
alignment of all proteins is produced and stored. The
distributions of duplicated and all paralogous pairs
shown in Fig. 3 are generated by averaging over 20
such samples.
\subsection*{Identification of true duplicated (sibling) pairs
by the Minimum Spanning Tree algorithm}

We are naturally not in possession of the set of
protein pairs that actually underwent duplication in
the course of evolution of a given genome.
The identification of the most likely candidates
for these ``true'' duplicates is in general a
rather complicated task which involves reconstructing the actual
phylogenetic tree for every family in a
genome. However, we
could make a much simpler educated guess about
past duplication events by connecting paralogous proteins in a given family
with the Minimum Spanning Tree (MST) that is
the tree maximizing the sum of PIDs along its edges
(or, to agree with its name, minimizing its opposite sign value). For
a family consisting of $F$  proteins such tree has
exactly $F-1$ edges representing our best guess about the actual
duplication events.
One can prove the truth of this by induction: when
a freshly duplicated pair is created
with PID=100\% it extends the
previously existing Minimum Spanning Tree of a family
by one edge. Assuming a constant rate of divergence
for all paralogous pairs in a given family, the set of
duplicated pairs would continue to form the Minimum
Spanning Tree at all times. We used the Kruskal
algorithm \cite{Kruskal56} to approximately detect the
MST.
\subsection*{Detection of families of paralogous
genes}

Families of paralogous proteins used in Figure 5 are
defined as mutually isolated clusters of proteins in the network in
which paralogous pairs are connected by a link.
Every two nodes within a family are either directly or indirectly
connected to each other by at least one chain of
paralogous links, while different clusters (families)
are completely disconnected from each other. Because
of our requirement for the length of the aligned
region to be $>$80\% of the length of the longest protein
in a pair, all proteins within such families are
rather homogeneous in their lengths.
%
\section*{Authors' contributions}
SM and designed the study and its analytical
framework. KKY acquired the genomic data and
performed the sequence alignment.
SM and KKY analyzed the results.
JBA and SM wrote the manuscript.
JBA performed numerical simulations of the
model proteome and the MST algorithm.
All authors read and approved the manuscript.
%
\section*{Reviewers' comments}
\subsection*{Reviewer 1: Eugene V Koonin, National Center
for Biotechnology Information, National Institute of Health, Bethesda, Maryland, USA}

This is quite an interesting, elegant study that
presents a mathematical model connecting the
distribution of percent sequence identity in
paralogous protein families with the parameter of the
gamma-distribution of intra-protein variability. The
latter parameter had been explored before, and the
values reported here are within the previously
estimated ranges, but to my knowledge, this is the
first work that derives this parameter theoretically
from completely independent data. It is intriguing
and, I suppose, important that the distributions of
the identities between paralogs and, accordingly, the
gamma-distribution parameter are almost
genome-independent. It seems like the latter parameter
is almost a "fundamental constant"  that follows from
the physics of protein structure that is, of course,
universal.

I have three comments that are rather technical but
bear on the robustness and  generality of the
conclusions.

1. A trivial point \ldots but, I feel it would have been
helpful to increase the number of analyzed genomes,
both in terms of diversity, and by including more than
one genome from each of the included lineages (and
others). The analysis of sets of related genomes would
(hopefully) demonstrate the robustness of the obtained
distributions, and would also help assessing the
significance of the differences in the exponents seen
among genomes. In particular, similar, flat
distributions found in human and in yeast are somewhat
strange given the huge difference in the size and
complexity of these genomes. This is attributed to the
legacy of whole-genome duplications but I find that
explanation dubious. Traces of this duplication in
yeast and, especially, in vertebrates are very weak.
Including more genomes would help to clarify this
issue. The reported genome analysis is very simple, it
cannot be computationally prohibitive.

{\bf Authors response:} {\it We agree that
extending our analysis to include more genomes is fairly straightforward.
However, we want to save the subject of lineage-dependence of the
exponent $\gamma$ (apart from that related to Whole Genome Duplications (WGD))
for future studies and to report it in a separate
publication.
To check our hypothesis that WGD are responsible for unusual profile of the PID-histograms
in human and yeast we analyzed the genome of {\it Paramecium
tetraurelia}. The lineage leading to this organism
underwent as many as four separately identifiable WGD
events. The results of our analysis presented in
Fig. 6 and the accompanying section of the manuscript have
beautifully confirmed our initial hypothesis:
while the all-to-all $N_a(p)$ histogram in the whole proteome of
{\it Paramecium tetraurelia} (solid diamonds in Fig. 6) has an unusually flat
profile similar to the one we saw in human (blue $\times$-es in Fig.
6), the removal of proteins generated in WGD
events results in a stepper PID-histogram (red stars in Fig. 6)
which is in excellent agreement
with the universal $p^{-4}$ functional form (dashed line in
Fig. 6). This provides necessary support to our
original conjecture that unusually flat PID histograms in human and baker's yeast
are also due to (possibly less obvious) duplicated pairs of proteins generated
in WGD events in these two lineages.
}

2. The mathematical model  developed in the paper is a
typical birth-and-death model. I wonder why the phrase
is not used (it would immediately clarify the matter
to those familiar with the field) and some of the
relevant  literature is not cited.

{\bf Authors response:} {\it We have modified our notation
to incorporate this comment. We also cited the appropriate
literature on birth-and-death models ([4-6])}

3. The ``true'' duplicates are identified using minimum
spanning tree under the constant rate assumption. This
is quite a crude method and an unrealistic assumption,
too. Building actual phylogenetic trees, certainly,
would be more appropriate. This might be too hard
technically for this amount of material but, at least,
the issues should be acknowledged, I think.

{\bf Authors response:} {\it In the Background section of the manuscript we
now explicitly mention that the minimum spanning tree is just a simpler (yet less precise)
alternative to reconstructing the actual phylogenetic tree for every family
in a genome.}

\subsection*{Reviewer 2: Yuri Wolf, National Center for Biotechnology Information,
National Institute of Health, Bethesda, Maryland, USA
(nominated by Eugene Koonin)}

The authors present an elegant model of protein
evolution that ties together duplication, loss of
paralogs and sequence divergence. Under the assumption
of the gamma-distributed variation of intra-protein
evolution rates the model correctly predicts the
power-law shape of the distribution of distances
between paralogs in fully sequenced genomes. Analysis
of the observed distributions allows to estimate the
long-term rates of duplication, retention of paralogs
and the shape parameter of the intra-protein evolution
rate distribution in different organisms.

The authors are very explicit and thorough about the
model description. However one point should be
emphasized for the sake of biologists among the
Biology Direct readership: the model is based on
neutral evolution of both protein sequence and the
complement of paralogs in the family and assumes that
all protein families behave in the same manner. This
is, obviously, a gross (if necessary) simplification
of reality. The good agreement between the model
predictions and the observed data is quite amazing and
possibly deserves some discussion.

{\bf Authors response:} {\it
I believe by this comment Dr. Wolf has raised an important point
which was inadequately presented in our manuscript.
While the model itself indeed was built using
a simplified (completely neutral) picture of real
evolutionary processes
the results of this analysis turned out to be independent
of these assumptions.
Thus they are expected to remain valid in a more
realistic evolutionary scenario (such as variable average
substitution, gene duplication and deletion rates in individual
families) when these assumptions
are relaxed. We have added a new subsection
``Robustness of the functional form of
$N_a(p)$ with respect to neutrality assumptions''
to the Results section of the manuscript which
describes in details our answer to this comment.
}

Probably the most interesting observation in the paper
is the near-constancy of the power of the middle part
of the distribution curve ($\sim$4) and, according to the
model, the shape parameter of the intra-protein
evolution rate distribution ($\sim$1/3). This result is in
surprising agreement with earlier estimates, including
our own [Grishin et al. 2000], obtained using entirely
different approaches. This might be telling us that
this parameter is a "universal constant" of protein
evolution and that it is dictated more by protein
physics rather than organism-specific properties.

{\bf Authors response:} {\it
This is now also discussed in the new section mentioned
in our previous response.
}

The section on the numerical simulations is somewhat
less justified in the eyes of this reviewer. The
underlying mathematical model appears to be fully
solved analytically and simulations follow the model
precisely. Thus the results of the simulations are
expected to agree with the analytical solution unless
some really dumb mistake was made in the course of
derivation or in the implementation of the
simulations. If I am missing something and the impact
of this section goes beyond the simple verification,
it probably should be discussed in the text.

{\bf Authors response:} {\it
We agree that for the most part we use numerical
simulations just to confirm the validity of our analytical
results. Still, we decided to keep it in the manuscript
since it clearly demonstrates (see Fig. 3) that the
deletion rate used in the model could be successfully
reconstructed from the $N_d(p)$ histogram. This is
not entirely obvious because of the approximations
that went into deducing the true duplicated (sibling) pairs
using the Minimum Spanning Tree algorithm.
}

\subsection*{Reviewer 3: David Krakauer, Santa Fe Institute, Santa Fe, New Mexico, USA}

The paper introduces a number of new ideas, including
a minimal spanning tree algorithm for eliminating
redundant distance information in a full pairwise
distance matrix in order to yield an estimate of the
true number of paralogous genes.

I tend to view this paper as a contribution to the
neutrality literature which seeks to explain large
scale patterns of genomic evolution in terms of
fundamental mutational processes without invoking
dedicated selection pressures acting on specific
genes. Having said this, selection could be playing an
important role in accounting for effective rate
variation in amino acid substitutions, and in
establishing the parameters of duplication and
deletion that prevent excessive growth or shrinking of
genomes. Whatever these selection pressures might be,
they would seem to have to apply across a number of
species.

I have a number of questions relating to the means of
establishing the empirical power law, and the
interpretation of the results.

1.  As the authors are aware straight lines in log-log
plots are not equivalent to having demonstrated a
power law distribution and least squares fitting
frequently generate biased estimates. Recent research
(Clauset et al 2007) presents maximum likelihood
estimators for  scaling parameters free from these
biases. How do the authors establish confidence in
their estimates of the exponent?

{\bf Authors response:} {\it
Due to an unavoidably narrow range of our power law
fits along the x-axis (The region II includes $0.3<p<0.95$
or half a decade) we didn't use any sophisticated
techniques in our fitting protocols. In fact, in this interval
the exponential fits look only marginally worse than those
with a power law. Thus for us the exponent
$\gamma$ of a power law fit is just a convenient single parameter
quantifying the distribution which is A) consistent
with the Gamma-distribution traditionally used
to describe substitution rates; B) nearly
universal in a broad variety of organisms ranging from
H. pylori to D. melanogaster.
}

2. I was somewhat confused by the renormalization
procedure for the raw distance histograms. While I
understand that this is required in order to ensure a
stationary distribution, I do not see clearly what the
biological implications or assumptions of this step
are. Perhaps this could be clarified?

{\bf Authors response:} {\it
A standard mathematical approach to describing stationary probability
distributions in growing systems is to normalize the histogram
in question by the sum of its elements ($\sum _p
N_a(p)$ in our case). However, we noticed that the
total number of paralogous pairs $\sum_p N_a(p)$ detected by
all-to-all alignment of all protein sequences
grows at the same exponential rate
$2(r_{\mathrm{dup}}-r_{\mathrm{del}})$ as
the square of the number of genes -- $N_{\mathrm{genes}}^2$.
Moreover, in real genomes the relationship
$\sum_p N_a(p) \sim N_{\mathrm{genes}}^2$
holds very well (see the newly created Supplementary
Figure S2). Thus we decided to use the
total number of gene pairs $N_{\mathrm{genes}}(N_{\mathrm{genes}}-1)/2 \sim N_{\mathrm{genes}}^2$
(the theoretical upper bound to the number of
paralogous pairs) to normalize the $N_a(p)$. The
biological implication of this result is that
the fraction of paralogous pairs among all
gene pairs is roughly the same in all genomes
(as manifested by the collapse of normalized distributions
in Fig. 1B).

We have also added the new subsection ``An estimate of the number
of superfamilies in different genomes'' in which we
speculate that the $p^{-4}$ could be extended well
into the region III down to its theoretical minimum of
$5-8\%$. This could be (at least partially) accomplished if
in addition to sequence similarity one would use
structural similarity to define superfamilies of
paralogous proteins. These results indicate that
normalizing $N_a(p)$ by $N_{\mathrm{genes}}(N_{\mathrm{genes}}-1)/2$
is a remarkably close approximation to normalizing it
by the actual (presently unknown) number of paralogous
pairs contained in the genome (including the evolutionary relationships
missed by sequence-alignment algorithms).
}

3. I  worried a little about the uniqueness of the
Gamma-distribution in generating the scaling behavior
given the absence of a robust test for the scaling
exponent. How many different distributions have been
tested, and how robust is the result to departure from
the Gamma?

{\bf Authors response:} {\it
As we explained in our response to your question \#1, due
to an inherently narrow range of the Region II (half a
decade) we never state that the power law functional
form
resulting from Gamma-distributed
substitution rates $\mu$
is the unique way to mathematically
describe the $N_a(p)$ histogram in real genomes.
Thus we didn't test multiple
$\mu$-distributions.
Our only claim is that the observed shape
of $N_a(p)$ is consistent with Gamma-distributed
($\alpha=0.33$) substitution rates of individual amino-acids within a protein.
}

4. A little more could have gone into the discussion
on the mutational processes and the role of selection.
Should we assume rate variation to be the outcome of
selection (as in the example of the slow rate of
evolution of ribosomal proteins) where perhaps an
active site remains more highly conserved, or is it
the contention of the authors that some purely
stochastic process at the mutational level accounts
for this variation? A similar argument could be made
for the duplication and deletion equilibrium. As the
paper reads now, I am not sure what processes the
authors have in mind.

{\bf Authors response:} {\it
These and other points are now explained in the new
subsection
``Robustness of the functional form of
$N_a(p)$ with respect to assumptions used in the model''
added to the Results section of the manuscript.
}

5. Surely the study of evolution through the
properties of genetic distance histograms is older
than Lynch and Conery (2000). I am aware of earlier
work by Gillespie and many others.

{\bf Authors response:} {\it
Thank you for pointing this out. We have added the book
by Gillespie and references therein to our citation
list.
}

6. The paper needs to be spell checked and grammar checked.

{\bf Authors response:} {\it
Done.
}

\subsection*{Reviewer 4: Eugene Shakhnovich, Harvard University, Cambridge, Massachusetts, USA}

Power-law distributions are ubiquitous in Protein Universe and were reported to describe
distribution of sizes of gene families, fold families, structural similarity relationships and
other properties (1-4). It had been widely accepted that the underlying reason for their
emergence is in evolutionary dynamics of creation of new genes and proteins and several
dynamics models have been proposed to describe it (1, 3, 5, 6).
Here the authors study the distribution of amino acid sequence similarities in
paralogous families and also find a regime where power-law describes the observed
histogram well. They proposed a mathematical model akin to master equation approach
which essentially assumes that the rate of divergence is non-uniform – it depends on the
sequence ID of genes in question.
The model and the analysis are interesting and make a valuable contribution to the
literature on evolutionary dynamics. The major strength of this study is in its highly
quantitative character which provides interesting insights about duplication/deletion rates
which are apparently dependent on past history. However I would like the authors to
address the following questions:

1) The authors consider all gene families in various organisms regardless of their
functional distributions. However recent work of B.Shakhnovich and Koonin (7)
(which is in a sense a forerunner of the present paper) has demonstrated that
evolution of paralogous families is dramatically different in the case of families
containing essential (E-families) genes and families that do not contain such
genes (N-families) . It would be interesting to carry out the same quantitative
analysis separately for E- and N-families and check how different exponents of
intermediate power-law regimes are and how does it fit into ID-dependent
divergence rate picture.

{\bf Authors response:} {\it
We agree that it would be interesting
to separately analyze the PID-histogram E- and N-families. The
Fig 3B of the B. Shakhnovich and E. Koonin article (Ref. 7 in the list below)
essentially does that.
From it one can see that while E-families, which are on average
larger (7) than N-families, have a sequence identity
histogram similar to the whole-genome $N_a(p)$ in our study,
the composite  PID histogram of all N-families in yeast
is essentially flat.
However, as we now explain in the revised version of our manuscript
our birth-and-death model does not apply to collections of
individual families grouped together
by some shared characteristic (e.g by their size or by whether
or not they contain an essential gene).
Indeed, the dynamics of such a group
would depend on additional parameters such as the rate
of creation and removal of families of a given type.
For example, the appearance of an essential gene
gene in an N-family would turn it into a new
E-family and remove its contribution to the
histogram of all N-families.
We feel that modification of our analysis to incorporate
these extra terms goes beyond the scope of this article.
%
%
%
%
%
%
}

2) The major weakness of this analysis (and other phenomenological approaches) is
that the ``explanation'' for power-law regime comes from an assumption of a
certain form of the distribution of substitution rates in the form of Gamma-function.
While assuming Gamma-function may result in good fits it is entirely
mysterious why does it emerge. The authors make a very interesting hint that
Gamma emerges from intra-protein variability of substitution rates but they do
not dwell much further on that. In fact such variability does exist. It was
quantitatively studied in a microscopic protein evolution model by Dokholyan and
myself in 2001 (8). It would be highly instructive to check whether distributions
of substitution rates observed in (8) can provide additional insights into the
microscopic origin of empirical fits used in this work.

{\bf Authors response:} {\it
It would be indeed extremely
exciting to find a truly ``microscopic''
explanation of the universal parameters of the Gamma-distribution
reported in our manuscript along the lines of the Ref.
(8). We feel however, that this goes well
beyond the scope of this article. In the new subsection
of our manuscript ``Robustness of the functional form of
$N_a(p)$ with respect to assumptions used in the model''
we now cite the Ref. (8) and mention that its results
lead to a biophysical explanation to
the remarkable universality of the exponent
$\alpha$ reported in our manuscript.
}

3) The effective distribution of intra-protein variability seems to depend on family
size. Why? Can it be related to functional constrains (e.g. E- and N- families). A
comment or further analysis will be helpful.

{\bf Authors response:} {\it
Soon after sending our manuscript for review we
realized that our mathematical model is
applicable only to whole genomes
or large individual families and
does not describe PID
histograms in collections of many families grouped by their
size. Indeed, such collections would have additional
birth-and-death events due to whole families
entering or leaving the selected bin of family sizes.
The rates of these processes would have a non-trivial
dependence on the age of a family and thus cannot be
easily incorporated into our mathematical framework.
Thus we removed the Fig. 5B showing the apparent systematic variation of the
exponent $\gamma$ with the family size, which, in the hindsight, is likely
caused by these extra birth-and-death terms.
For the discussion of E- and N-families see our response
to your question 2).
}

4) The power-law regime is observed only at sufficient level of divergence. Why?
How can current model be modified to account for the full histogram, not only its
power-law part?

{\bf Authors response:} {\it
We attribute the upward turn in $N_a(p)$ for $p>90$\% (inside the Region I)
to much higher deletion rates of recently duplicated genes caused by their
apparent redundancy. The crossover to power law
in the Region II means that this redundancy tends to be
lost below this level of sequence identity. The combination of
our equations (1) and (7) provide a comprehensive mathematical description
valid in both Regions II and I (as we explained the crossover in the region III is an artifact
caused by some bona fide paralogous pairs being missed by sequence alignment algorithms).
}

{\bf References used by Eugene Shakhnovich:}
1. Qian, J., Luscombe, N. M. \& Gerstein, M. (2001) J Mol Biol 313,
673-81.\\
2. Koonin, E. V., Wolf, Y. I. \& Karev, G. P. (2002) Nature 420,
218-23.\\
3. Huynen, M. A. \& van Nimwegen, E. (1998) Mol Biol Evol 15,
583-9.\\
4. Dokholyan, N. V., Shakhnovich, B. \& Shakhnovich, E. I. (2002) Proc Natl Acad
Sci U S A 99, 14132-6.\\
5. Karev, G. P., Wolf, Y. I., Rzhetsky, A. Y., Berezovskaya, F. S. \& Koonin, E. V.
(2002) BMC Evol Biol 2, 18.\\
6. Roland, C. B. \& Shakhnovich, E. I. (2007) Biophys J 92,
701-16.\\
7. Shakhnovich, B. E. \& Koonin, E. V. (2006) Genome Res 16,
1529-36.\\
8. Dokholyan, N. V. \& Shakhnovich, E. I. (2001) J Mol Biol 312, 289-307.
%
\section*{Acknowledgements}
Work at Brookhaven National Laboratory was carried out under Contract
No. DE-AC02-98CH10886, Division of Material Science, U.S. Department
of Energy. JBA thanks the Institute for Strongly Correlated and
Complex Systems at Brookhaven National Laboratory for hospitality and
financial support during visits when the majority of this work was
done. JBA thanks the Lundbeck Foundation for sponsorship of PhD
studies at the Niels Bohr Institute. JBA and SM
visit to Kavli Institute for Theoretical Physics where this
work was initiated was supported by the National Science
Foundation under Grant No. PHY05-51164.
%
%
%
%
%
%
%
%
%
%
%
%
%

{\ifthenelse{\boolean{publ}}{\footnotesize}{\small}
 \bibliographystyle{bmc_article}  
  \bibliography{refs} }     

\ifthenelse{\boolean{publ}}{\end{multicols}}{}

\section*{Figures}

  \subsection*{Figure 1 - Histogram of all amino-acid sequence identities}
The histogram $N_a(p)$ (panel A) and the normalized
histogram $n_a(p)=N_a(p)/[N_{\mathrm{genes}}(N_{\mathrm{genes}}-1)/2]$ (panel B)
of amino acid sequence identities
$p$ for all pairs of paralogous proteins in complete genomes of {\it
H. pylori} (blue stars), {\it E. coli} (green open squares), {\it
S. cerevisiae} (red crosses), {\it C. elegans} (cyan open triangles),
{\it D. melanogaster} (magenta filled circles) and {\it H. sapiens}
(brown filled triangles). The dashed line is a power-law
$p^{-4}$. Note the logarithmic scale of both axes.  Vertical lines
separate regions I, II and III described in the text.
%
  \subsection*{Figure 2 - Illustration of dynamical processes changing the
  PID histogram.}

(panel A) The PID decay generates a
negative flux $v(p)N_a(p)$ down the PID-axis. The net flux into a
given bin $\Delta p$ is given by $v(p+\Delta p)N_a(p+\Delta
p)-v(p)N_a(p) \approx
\Delta p \frac{d}{dp}[v(p)N_a(p)]$
(panel B) A single gene duplication event $A \to A'$ gives rise to
three new paralogous pairs: $A'-A$, $A'-B$ and $A'-C$.
Immediately after the duplication the pair $A-A'$
has the PID=100\% , while PIDs of $A'-B$ and $A'-C$
are equal to those of $A-B$ and $A-C$.
Thus the PID of every previously existing paralogous
pair involving $A$ gets duplicated along with the duplication
$A \to A'$.
  \subsection*{Figure 3 - Histogram of sequence identities in a numerical simulation
  of proteome dynamics.}

The histogram of sequences identities of all
paralogous pairs $N_a(p)$ (filled circles) and duplicated pairs
$N_d(p)$ (open circles) in an artificial proteome generated by our
numerical model for $\alpha =0.33$ and $r_{dup}/\bar{\mu}=r_{\mathrm{del}}/\bar{\mu} =
0.25$ as described in the text. Solid line has the slope -4, while
the dashed one is given by
\ref{eq:Ndfinal} with $\gamma=4$ and the best fit
value of $r_{\mathrm{del}}/\bar{\mu} =0.237$.
%
%
  \subsection*{Figure 4 - Correlation between the number of genes in an organism and its
  duplication/deletion rates.}

Evolutionary parameters
$r_{\mathrm{dup}}/\bar{\mu}$ (open diamonds),
$r_{\mathrm{del}}/\bar{\mu}$ (filled circles),
$r_{\mathrm{dup}}^*/\bar{\mu}$ (open squares),
and
$r_{\mathrm{del}}^*/\bar{\mu}$ (filled triangles)
plotted versus the total
number of genes $N_{\mathrm{genes}}$ in an organism.
Organisms in the order of increasing number of genes are
{\it H. pylori, E. coli},
{\it S. cerevisiae},
{\it D. melanogaster, C. elegans}, and {\it H. sapiens}.
As explained in the text, more complex
organisms (those with larger $N_{\mathrm{genes}}$) tend to be
characterized by higher values of the first three ratios
but lower values of the last ratio.
   \subsection*{Figure 5 - Histogram of sequence identities of individual
families in the genome of {\it C. elegans}.}

The histogram   of amino acid sequence identities
for pairs of paralogous proteins contained in each of
the 5 largest families in the genome of {\it C. elegans}.
Families in the
order of decreasing size (measured by the number of
proteins) are marked with green stars (243 proteins),
cyan squares (188 proteins),
red x's (162 proteins), brown triangles (105 proteins), and magenta +'s
(73 proteins). Solid blue
circles show the distribution of all paralogous pairs
in the genome (as in Figure 1), while solid black
diamonds - what is left after removing the above 5
largest families. Dashed line corresponds to a power
law with the slope -4, while the solid one - the slope -5.
%

   \subsection*{Figure 6 - Histogram $N_a(p)$ of sequence identities and
four rounds Whole Genome Duplications (WGD)
in {\it Paramecium tetraurelia}.}
{
The histogram of sequence identities of 103,828 paralogous pairs
among 39,642 proteins in the genome of {\it Paramecium tetraurelia}
(solid diamonds) detected by an all-to-all BLASTp alignment
(see methods for details). Other histograms shown in this plot
correspond to a progressive
removal of new proteins created in the four WGD events
\cite{ciliate_WGD}:
41,890 pairs among  27,616 proteins
excluding those generated in the latest
WGD event (solid squares), 25,342 pairs among 23,618 proteins
excluding those generated in the last two WGD events (solid circles),
22,287 pairs excluding those generated in the latest
three WGD events (open triangles), and 21,417 pairs
among 22,635 proteins excluding those generated in
all four known WGD events (red
stars). For comparison we copy from Fig. 1A
the histogram of 31,078 pairs among 25,319 {\it H.
sapiens} proteins (blue $times$-es).
One can see that by progressive elimination
of pairs generated in WGD events the functional form
of the $N_a(p)$ histogram in {\it Paramecium tetraurelia}
approaches the universal scaling form:
$N_a(p) \sim p^{-4}$ (dashed line).
}
\section*{Tables}

  \subsection*{Table 1 - Statistics of datasets used in this study.}

The first column is the name of the organism, the second column
-- the number of protein-coding genes in its genome, $N_{\mathrm{genes}}$,
the third column - the number of proteins for which we found
at least one paralogous partner, the fourth column is the percentage
of proteins with at least one paralog,
the fifth column - the total number
of distinct BLAST hits generated before we applied subsequent filtering,
the sixth column - the number of paralogous pairs included in $N_a(p)$,
and the seventh column - in $N_d(p)$. \par \mbox{}
\par
\mbox{
\begin{tabular}{l | r | r | r | r | r | r}
{\bf Organism}
&
\begin{tabular}[t]{p{1.2cm}}
Proteome\\ size
\end{tabular}
&
\begin{tabular}[t]{p{1.5cm}}
Number of \\ proteins\\ with \\ paralogs
\end{tabular}
&
\begin{tabular}[t]{p{1.5cm}}
\% of \\ proteins\\ with \\ paralogs
\end{tabular}
&
\begin{tabular}[t]{p{1.5cm}}
BLASTP\\ hits
\end{tabular}
&
\begin{tabular}[t]{p{1.5cm}}
Number \\of pairs\\ in $N_a(p)$
\end{tabular}
&
\begin{tabular}[t]{p{1.5cm}}
Number \\of pairs\\ in $N_d(p)$
\end{tabular}\\
\hline \hline
{\it H.pylori}       & 1590  & 230  &  14\% & 3228   &   260 &  148 \\
{\it E.coli}         & 4288  & 1428 &  33\% & 16768  &  2614 & 1013 \\
{\it S.cerevisiae}   & 5885  & 1689 &  29\% & 43915  &  2297 & 1025 \\
{\it C.elegans}      & 19099 & 6894 &  36\% & 204398 & 46463 & 5545 \\
{\it D.melanogaster} & 14015 & 4153 &  30\% & 557047 & 17621 & 3238 \\
{\it H.sapiens}      & 25319 & 9252 &  37\% & 1330721 & 31078 & 6595 \\
\hline
\end{tabular}
}
 \subsection*{Table 2 - Deletion and duplication rates }

The first column contains the name of the organism, the
second column --  $N_{\mathrm{genes}}$, the number of genes in its genome,
the third column is the value of the exponent $\gamma$ in
the best fit with $p^{-\gamma}$ to $N_a(p)$ in the region II.  The fourth,
fifth, sixth and seventh columns are correspondingly the ratios
$r_{\mathrm{dup}}^*/\bar{\mu}$,
$r_{\mathrm{dup}}/\bar{\mu}$, $r_{\mathrm{del}}/\bar{\mu}$,
and $r_{\mathrm{del}}^*/\bar{\mu}$ defined and measured as
described in the text. \par \mbox{}
\par
\mbox{
\begin{tabular}{l | p{2.5cm} | p{1cm}| p{1.5cm}| p{1.5cm}| p{1.5cm}| p{1.5cm}}
{\bf Organism} & Proteome size & $\gamma$ & $r_{\mathrm{dup}}^*/\bar{\mu}$
& $r_{\mathrm{dup}}/\bar{\mu}$ & $r_{\mathrm{del}}/\bar{\mu}$ &
$r_{\mathrm{del}}^*/\bar{\mu}$\\
\hline \hline
{\it H.pylori}       &  1590 &   3.1 &   0.73 &  0.032 &    0.16 & 67 \\
{\it E.coli}         &  4288 &   4.4 &   1.37 &  0.038 &    0.10 & 64  \\
{\it S.cerevisiae}   &  5885 &   1.8 &   1.61 &  0.24 &    0.24 & 27 \\
{\it C.elegans}      & 19099 &   4.2 &   3.16 &  0.27 &    0.37 & 41 \\
{\it D.melanogaster} & 14015 &   4.4 &   0.35 &  0.084 &    0.22 & 30  \\
{\it H.sapiens}      & 25319 &   2.4 &   2.82 &  0.85 &    0.16 & 19 \\
\hline
\end{tabular}
}
\section*{Additional Files}
\subsection*{Additional figure S1 ---
The overall shape of the PID histogram is independent of the alignment algorithm and the E-value
cutoff.}

The PID histogram $N_a(p)$ in the fly ({\it D. melanogaster}
genomes when pairs of
paralogous proteins were detected using the blastp
algorithm \cite{Altschul90} with E-value cutoff
of $10^{-10}$  (filled circles) and $10^{-30}$  (open diamonds).
The inset shows the ratio of these two histograms, which is very
close to 1 for $p>40\%$. Thus the overall shape of $N_a(p)$
in most of the Region II (Fig. 1) is nearly +cutoff independent.
The $N_a(p)$ also is
insensitive to a particular algorithm used to align
the pairs. Indeed, when paralogous pairs
detected by the blastp with the E-value cutoff of $10^{-10}$
(filled circles) were realigned using the
Smith-Waterman algorithm \cite{SmithWaterman81} the
resulting distribution (blue stars) changed
very little.

\subsection*{Additional figure S2 ---
The quadratic scaling of the total number of paralogous pairs with the
number of genes in the genome.}
The total number of paralogous pairs $\sum_p N_a(p)$ generated by
the all-to-all alignment of all protein sequences
encoded in the genome (the y-axis) scales as the
square of the total number  $N_{\mathrm{genes}}$ of protein-coding
genes in the genome. Solid symbols are six model organisms used in our study.
The solid line has the slope 2 on this log-log plot.

\end{bmcformat}

\includegraphics[width=\textwidth]{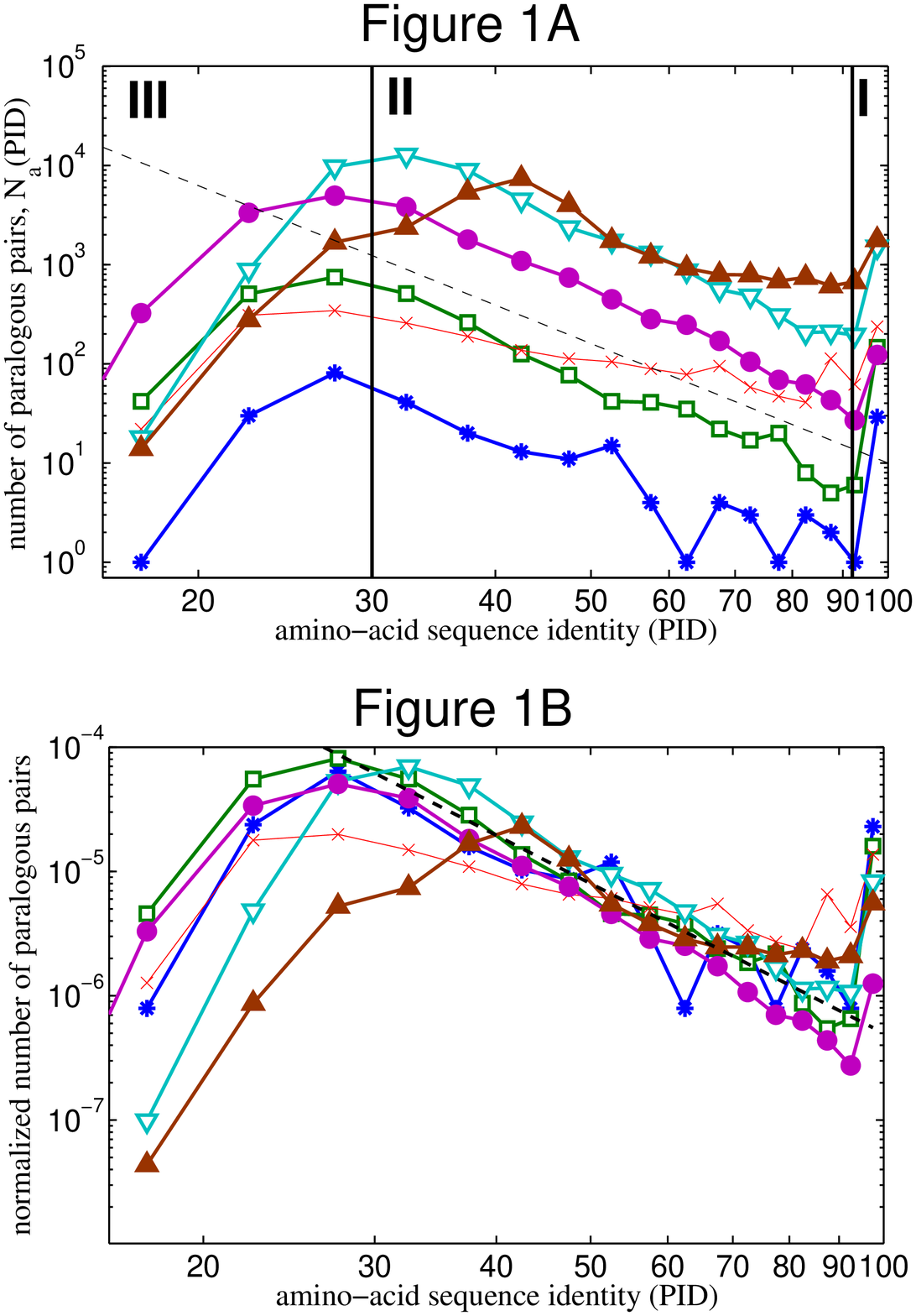}\\
\includegraphics[width=\textwidth]{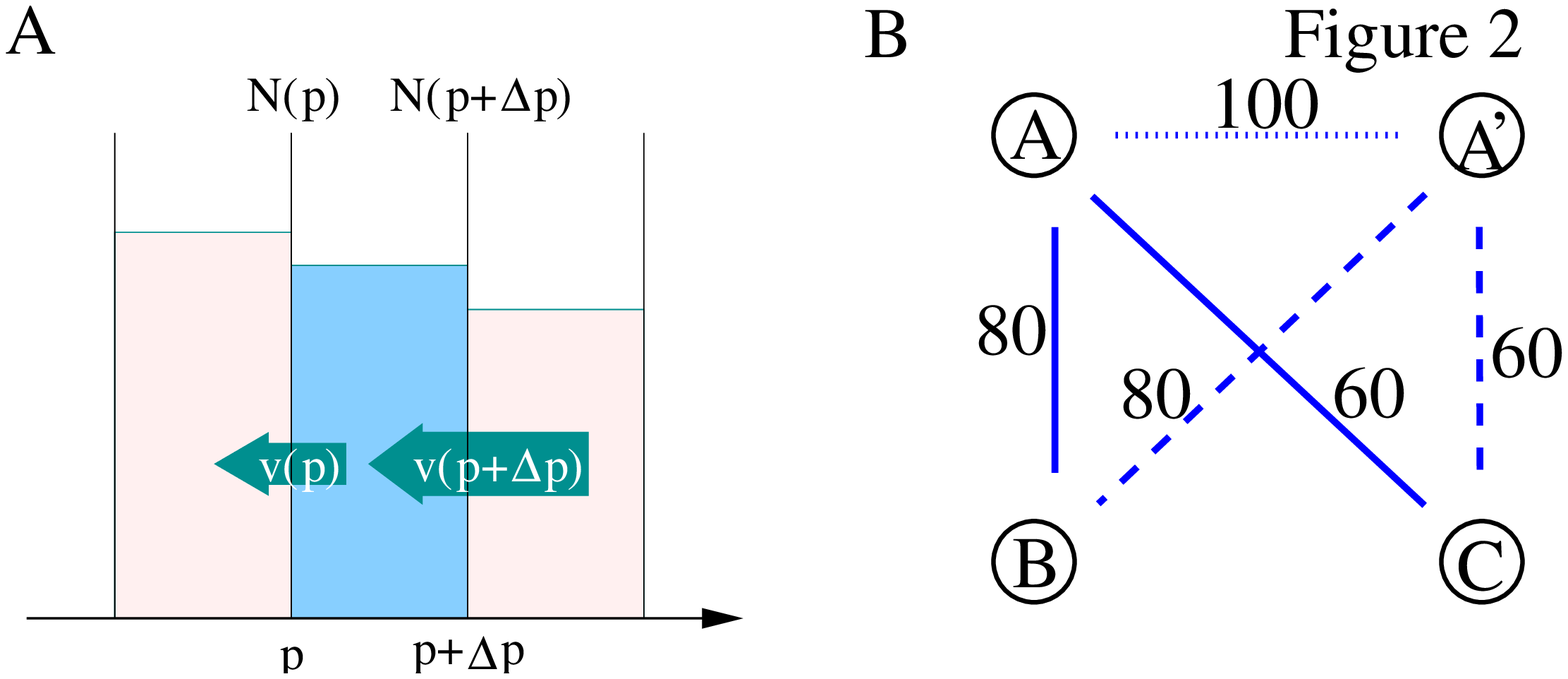}
\includegraphics[width=\textwidth]{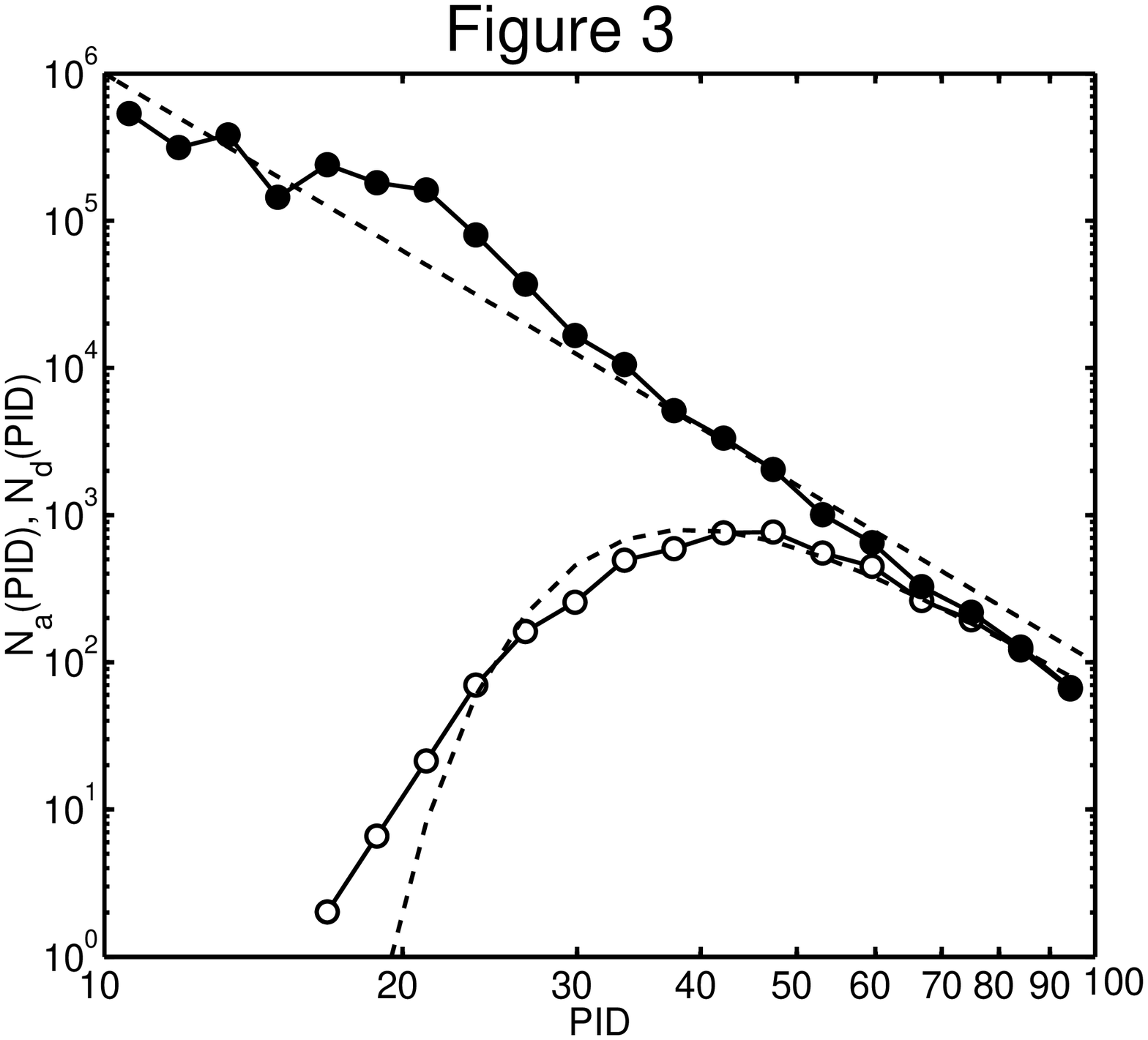}\\
\includegraphics[width=\textwidth]{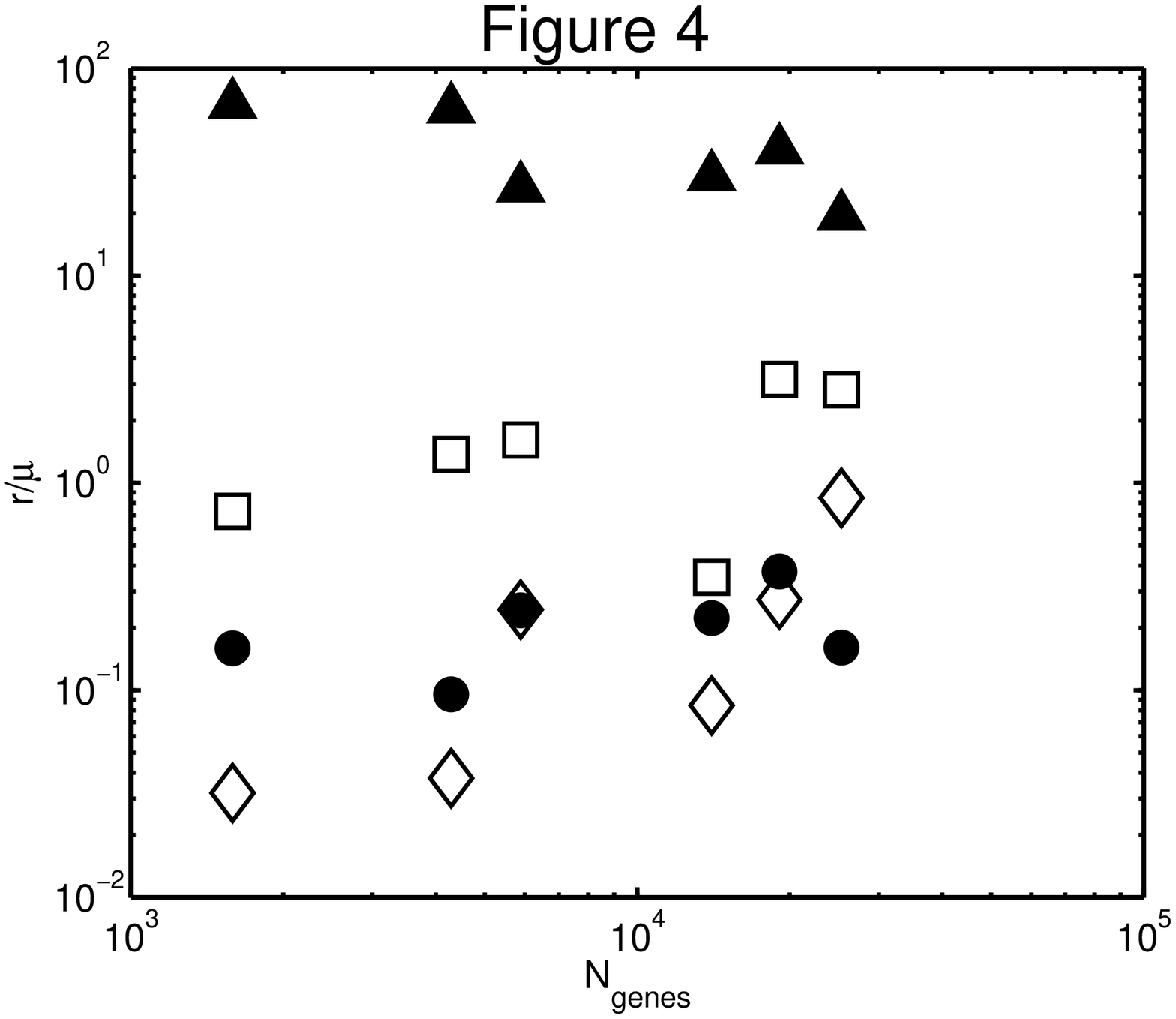}\\
\includegraphics[width=\textwidth]{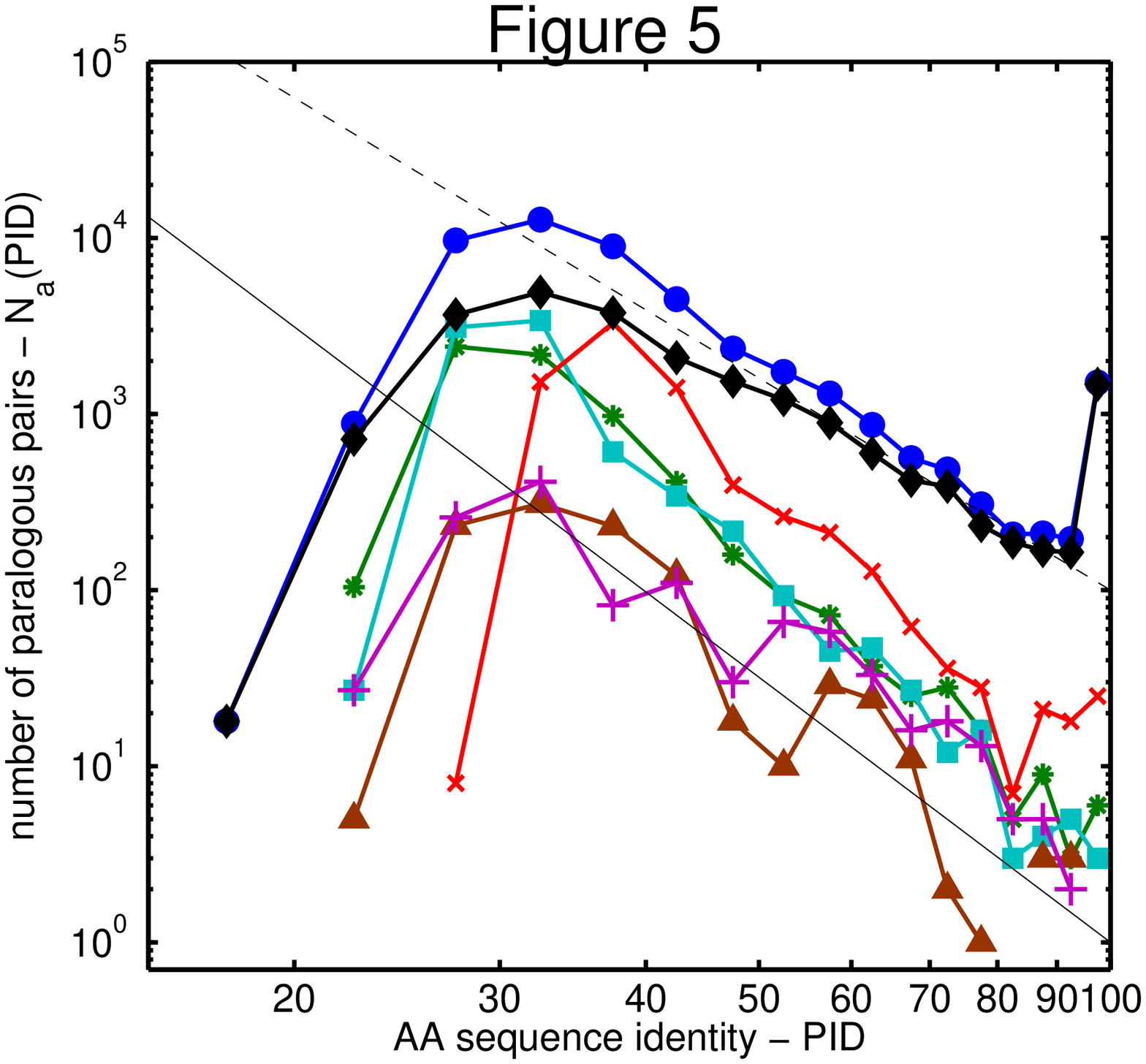}\\
\includegraphics[width=\textwidth]{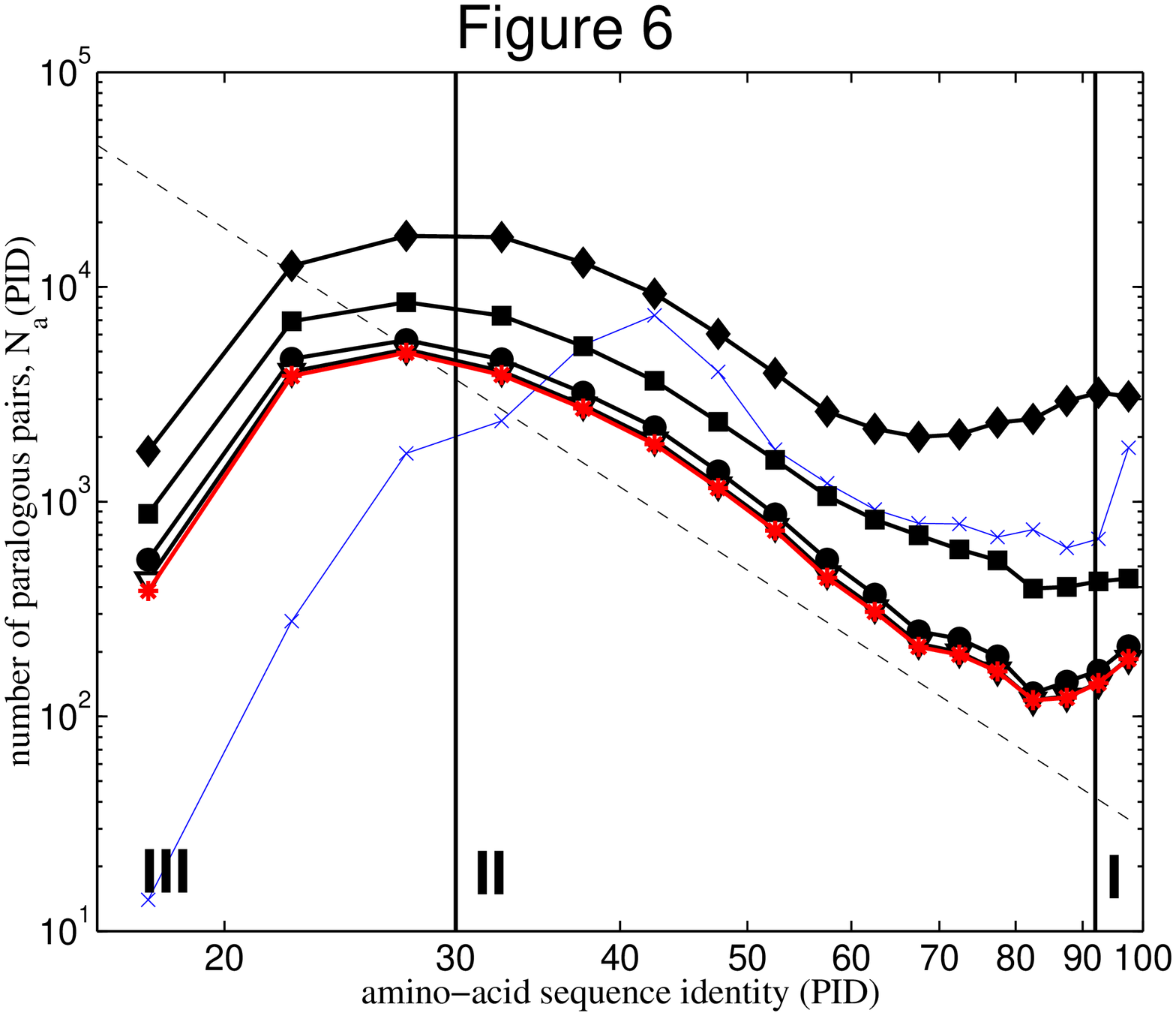}\\
\includegraphics[width=\textwidth]{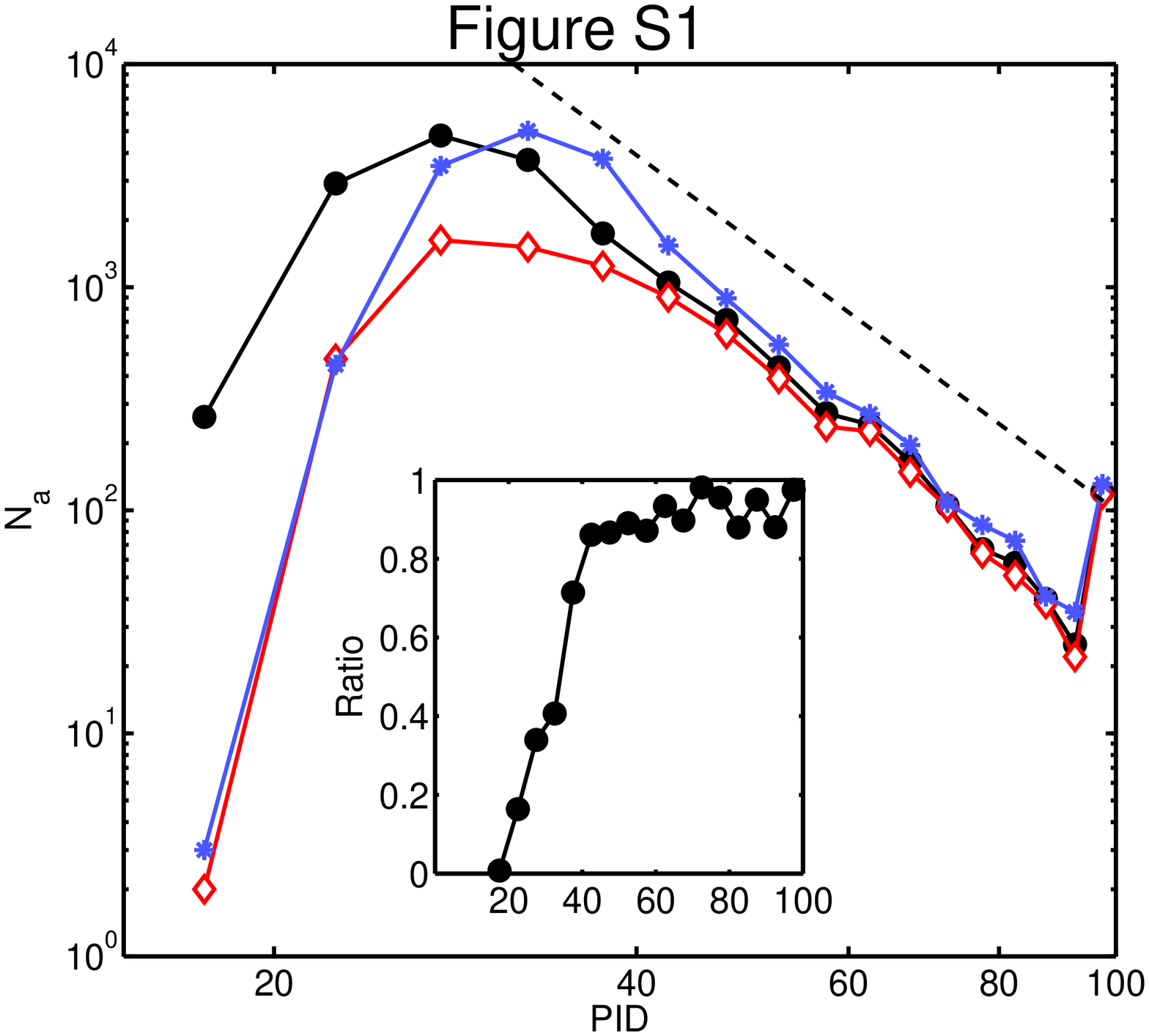}\\
\includegraphics[width=\textwidth]{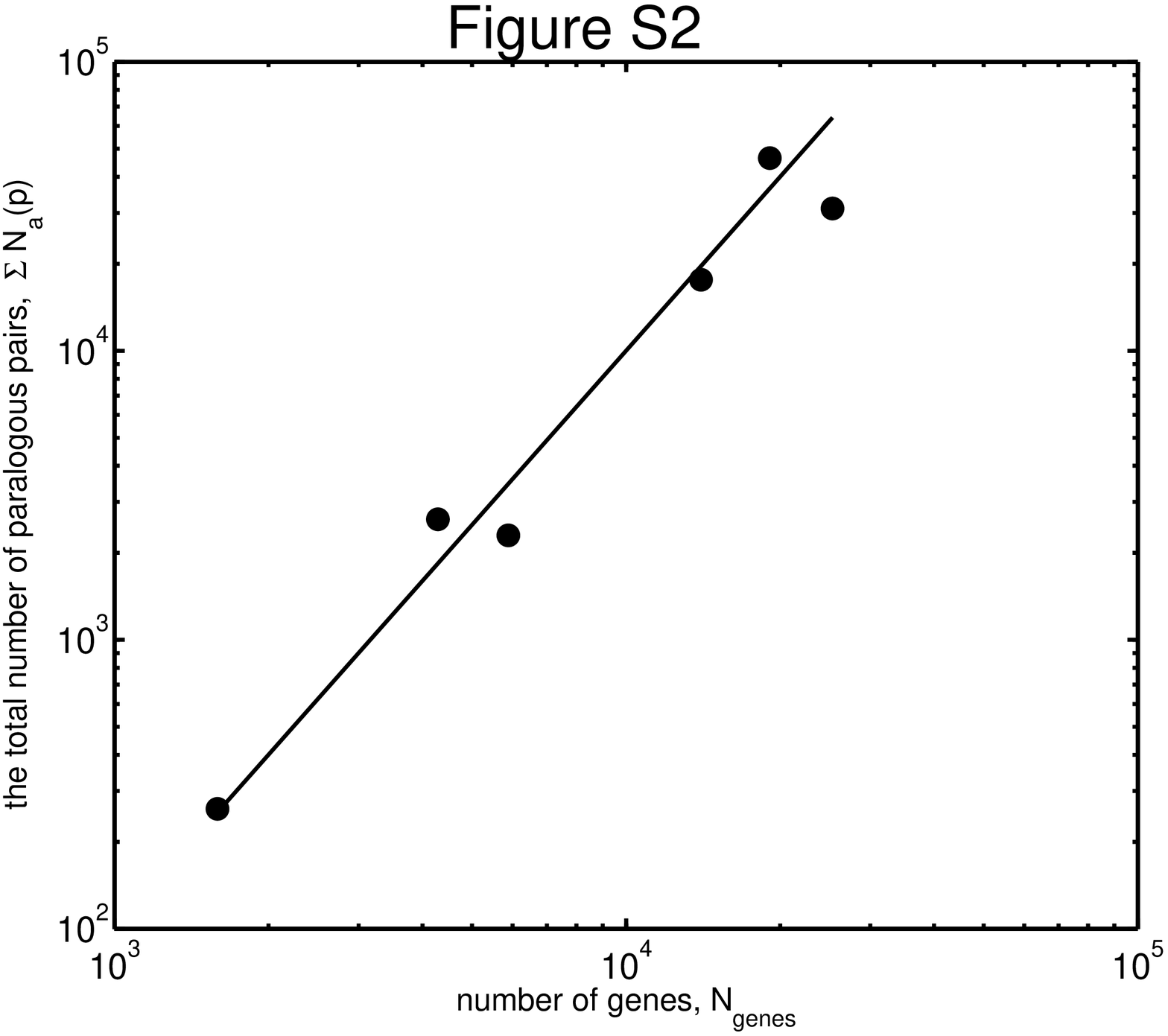}

\end{document}